\documentclass{aa}  

\usepackage{graphicx,subfigure}
\usepackage{txfonts}
\usepackage[unicode]{hyperref}
\begin{document}

\title{A comprehensive X-ray analysis of the massive O-type binary \\HD\,93250 over two decades}

   \author{Bharti Arora
          \inst{1}, Micha\"{e}l De Becker\inst{1} \and Jeewan C. Pandey\inst{2}
          }

   \institute{Space Sciences, Technologies and Astrophysics Research (STAR) Institute, University of Li\`ege, Quartier Agora, 19c, All\'ee du 6 A\^out, B5c, B-4000 Sart Tilman, Belgium\\ 
              \email{bhartiarora612@gmail.com}
             \and
             Aryabhatta Research Institute of Observational Sciences (ARIES), Nainital$-$263 001, India
             }

   \date{Received...; accepted...}
\authorrunning{B. Arora et al.}
\titlerunning{Comprehensive analysis of X-ray emission from HD\,93250}

 
  \abstract
   {Massive star winds are known to be responsible for X-ray emission arising from wind plasma heated by the strong shocks up to the temperature of 10$^6$--10$^7$ K in case of colliding wind binaries. The investigation of X-ray emission from massive stars thus constitutes a valuable tool to identify binaries which otherwise is difficult to determine using classical techniques.}
   {We have investigated thermal and non-thermal X-ray emission from the massive O-type star HD\,93250 to unveil its binary orbital parameters independently.}
   {To meet our goal, X-ray data obtained with European Photon Imaging Camera onboard \textit{XMM-Newton} has been analyzed, spanning over $\sim$19 years.  Additionally, we analyzed \textit{NuSTAR} observations of HD\,93250 taken at various epochs.}
   {We determined the variability time-scale of the X-ray emission to be 193.8$\pm$1.3\,d, in full agreement with the 194.3$\pm$0.4\,d period derived from the astrometric orbit. The X-ray spectrum of HD\,93250 is well explained by a three-temperature thermal plasma emission model with temperatures of 0.26, 1.0, and 3.3 keV. The resulting X-ray flux varies in compliance with the typical colliding wind emission from eccentric massive binaries where it enhances near periastron passage and decreases gradually close to apastron, proportionally with the inverse of the binary separation. The periastron-to-apastron X-ray emission ratio points to an eccentricity range of 0.20-0.25, once again in agreement with the previously determined astrometric orbit. Finally, we did not detect any hard X-ray emission attributable to non-thermal emission above 10 keV.}
   {Given the derived plasma temperature, the strong phase-locked variability and the significant over-luminosity in X-rays, we establish that the X-ray emission from HD\,93250 is dominated by the colliding-wind region. Our results lend support to the idea that X-ray time analysis of massive stars constitutes a relevant tool to investigate their multiplicity and extract relevant information on their basic orbital parameters, such as the period and the eccentricity, independently of any orbital solution derived from usual techniques.}

   \maketitle
%

\section{Introduction}\label{intro}

The Carina nebula region is an interesting star-forming region of the galaxy containing some of the youngest and most massive O-type stars. HD\,93250 is an O-type member of the open cluster Trumpler 16, which was seen in X-rays first with the high-resolution imager (HRI) and imaging proportional counter (IPC) onboard \textit{Einstein} observatory in 1978 \citep{1979ApJ...234L..55S}.   

The star has been classified as of spectral type O3 V((f)) \citep{1971ApJ...167L..31W,1972AJ.....77..312W}. \citet{2002AJ....123.2754W} revised the spectral type of HD\,93250 to O3.5 V((f+)) and later to O4 III(fc) by \citet{2010ApJ...711L.143W} looking at its C III emission lines. It was again reclassified as O4 IV(fc) since it is similar to the other class IV object as observed by \citet{2016ApJS..224....4M}. \citet{2003ApJ...589..509E} discussed the probable binarity of the source from its strong and hard X-ray emission. Similarly, \citet{2008A&A...477..593A} studied 5 observation IDs from XMM-Newton spanned over a year and explained the X-ray spectra of HD\,93250 with multiple components of the optically thin thermal plasma model. Several spectroscopic and radial velocity measurements performed over a few decades have missed any clear evidence for the binary nature of the source \citep{1973MmRAS..77..199T,1982ApJ...254L..15W,1987ApJS...64..545G,1996ApJ...463..737P,2009MNRAS.398.1582R,2011AJ....142..146W}. However, \citet{2011ApJS..194....5G} suggested that HD\,93250 could be either an O+O binary with period $>$30 days or a magnetic star looking at its enhanced and variable X-ray emission with \textit{Chandra} observations. In parallel, \citet{2011ApJ...740L..43S} announced an interferometric detection of a similar binary companion at a separation of $\sim$1.5 mas, corresponding to 3.5 AU at the distance of Carina using the ESO Very Large Telescope Interferometer (VLTI) observation made with the instrument Astronomical Multi-BEam combineR (AMBER) on 2010 December 27. Later, \citet{2017A&A...601A..34L} constrained the astrometric orbit of HD\,93250 and derived the binary orbital period of 194.31$\pm$0.39 days, with an eccentricity of 0.217$\pm$0.011. They utilized data explored by \citet{2011ApJ...740L..43S} in addition to more interferometric observations obtained from AMBER and Precision Integrated-Optics Near-infrared Imaging ExpeRiment (PIONIER) instrument of VLTI for HD\,93250. The latter study suggested the system to consist of similar high mass O4 type components. A low angle of inclination as well as almost similar mass binary components, have been suggested to cause the non-detection of variations in its radial velocity during the past investigations. 

Our motivation to carry out the detailed study of HD\,92350 is to characterize the variations in its X-ray emission according to the findings of \citet{ 2017A&A...601A..34L} and \citet{2011ApJ...740L..43S}. Colliding winds have been speculated to significantly contribute to the X-ray emission from HD\,93250 by the studies mentioned earlier. In light of the clarification of the orbital properties of the system, it is now timely to reconsider the question of the X-ray emission from HD\,93250. With this aim, we have investigated X-ray observations of HD\,93250 over a time baseline with unprecedented duration for this object.  

This paper is organized as follows. Section \ref{intro} describes the target and summarizes the outcome of previous works relevant to our purpose. Section \ref{x-ray} presents the X-ray observations and the processing of data used for the present study. The X-ray light curve and spectral analysis are detailed in Sects. \ref{lc} and \ref{spec}, respectively. Finally, Sect. \ref{disc} includes a discussion of our main results, while conclusions are drawn in Sect. \ref{conc}.


\section{Observations and data reduction}\label{x-ray}

\subsection{X-ray monitoring of HD\,93250 with \textit{XMM-Newton} }
X-ray data of HD\,93250 obtained with \textit{XMM$-$Newton} \citep{2001A&A...365L...1J} from 2000 July to 2019 December has been analyzed. A total of 33 epochs of X-ray observations have been utilized as logged in Table \ref{tab:log}. All of the observations were obtained for other massive stars as the main target in the Eta Carinae field and HD\,93250 happened to lie in the same field of view of \textit{XMM-Newton}. It was observed with different configurations of the three European Photon Imaging Camera (EPIC) instruments, \textit{viz.} MOS1, MOS2, and PN. The MOS1 and MOS2 cameras took data mostly in the prime-full imaging mode, but a few observations were also done in the prime-partial window--W2$/$W3 imaging modes. However, the PN camera operated either in prime-full or prime-large window mode. The data reduction was performed by using the latest calibration files with SAS v20.0.0.

The raw EPIC Observation Data Files (ODF) were pipeline processed using the tasks \textsc{epchain} and \textsc{emchain} for the PN and MOS data, respectively. The SAS task \textsc{evselect} generated the list of event files by considering the good events having pattern 0$-$4 for PN and 0$-$12 for MOS data. The data was found to be unaffected by pile-up after examining with the task \textsc{epatplot}. To check the intervals of high background emission, full-frame light curves were generated considering the single-event (PATTERN = 0) in $>$10 keV energy range for MOS and that in the 10-12 keV band for PN. Good time intervals were selected by removing the intervals with abruptly high background. Typically, the filtering criterion excluded the intervals of count rate higher than 0.20 counts s$^{-1}$ for MOS and  0.4 counts s$^{-1}$ for PN background light curves.

\begin{figure}
\centering
    \includegraphics[width=0.93\columnwidth]{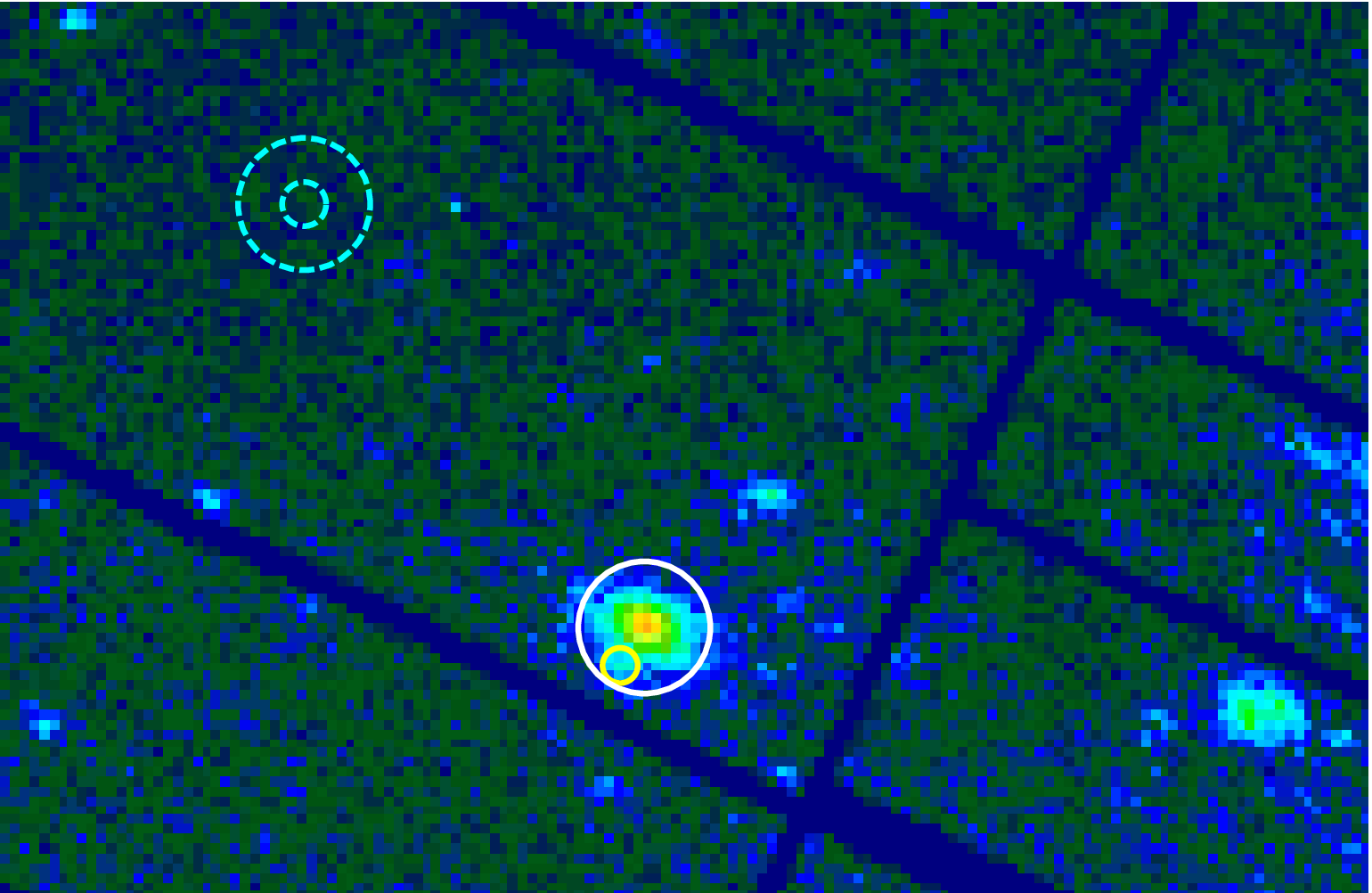}
	\caption{\textit{XMM-Netwon}$-$PN false-color image of HD\,93250 from observation ID 0311990101. The white circle represents the source region of a 30 arcsec radius at the location of HD\,93250. However, the X-ray counts from the neighboring source region of 10 arcsec radius (in yellow color) were subtracted from the target light curve and spectra. The background has been estimated from an annular region (cyan color) of 10 and 30 arcsec inner and outer radii in the source-free area of the detector.  \label{fig:fig1}}
\end{figure}

The PN image of HD 92350 obtained from observation ID 0311990101 in 0.3$-$12.0 keV energy range is shown in Fig. \ref{fig:fig1}. To estimate the source coordinates precisely, we made use of the standard source detection algorithm in the SAS through the meta-task \textsc{edetect\_chain} applied to PN data which has better photon statistics than MOS detectors. The target source HD\,93250 (S1) was detected at the position R.A. (J2000) = 10:44:45.1200 and Dec (J2000) = -59:33:52.920. Another source (S2) at position  R.A. (J2000) = 10:44:46.5600 and Dec (J2000) = -59:34:10.200 was detected closest to the source position as shown in Fig. \ref{fig:fig1}. Both of these sources were detected in \textit{Chandra} Carina Survey \citep{2011ApJS..194....2B}. \textit{Chandra} has better sensitivity to determine precise source positions and our estimates agree well with \textit{Chandra} detection. The source S2 has been identified as a B5III-V type star named Cl* Trumpler 14 Y 66 emitting in X-rays \citep{2015AJ....150..195V}. A circular region with a radius of 30 arcsec centered at the source coordinates was selected to accumulate enough counts for the extraction of EPIC light curves and spectra of HD\,93250. To remove the contamination from the S2 neighboring X-ray source, the counts from another circular region of 10 arcsec radius centered at the location of S2 were subtracted from the actual source region. This extraction radius accounts for an encircled energy fraction of about 60\,\%\footnote{\url http://xmm-tools.cosmos.esa.int/external/xmm\_user\_support/\\documentation/uhb/}. Among the remaining maximum 40\,\%, a fraction of more than one third is actually out of the S1 extraction region. Given the X-ray fluxes reported by \citet{2003ApJ...589..509E}, i.e. $2.3\times10^{-12}$ and $1.3\times10^{-13}$\,erg\,s$^{-1}$\,cm$^{-2}$ for S1 and S2, respectively, we estimate that at most about $1.3\times10^{-14}$\,erg\,s$^{-1}$\,cm$^{-2}$ should contaminate S1 extraction region. This corresponds to $\sim$1\% of the flux from our main target, confirming the very low contamination level. We also note that the counts measured in the S2 extraction region contributes $\sim$5\% of the total counts measured in the circular area around S1, in fair agreement with the {\it Chandra} values mentioned above.

Background estimation was done from an annular region with inner and outer radii of 10 and 30 arcsec, respectively, ensuring the same detector area for the background region as the source. The background region has been chosen as close as possible to the source in its surrounding source-free area to choose both the regions at nearly the same off-axis angle from the detector aim-point. It has also been made sure that no source is detected in the selected background region by \textit{Chandra} which has better resolution and sensitivity than \textit{XMM-Newton}--EPIC.  The obtained light curves were further corrected for good time intervals, dead time, exposure, point-spread function, and background subtraction using the \textsc{epiclccorr} task. The source as well as the background spectra were generated by the task \textsc{evselect}. The dedicated ARF and RMF response matrices required for calibrating the energy and flux axes were calculated by the tasks \textsc{arfgen} and \textsc{rmfgen}, respectively. The backscaling of the extracted spectra was done using the task \textsc{backscale}. In order to have a minimum of 15 counts per spectral bin, the EPIC spectra were grouped using \textsc{grppha}. Further temporal and spectral analyses were performed using HEASoft version v6.29c.

HD\,93250 was present in the field of view of MOS1 and MOS2 during most of the epochs of observation but was much less well exposed in the PN field of view due to CCD gaps of bad columns at several epochs. Therefore, we did not consider PN data any further to keep the current analysis consistent and coherent across the full-time series.
                                                                                             
\longtab{
\begin{longtable}{c c c c c c c c c c}
\caption{\label{tab:log} Log of \textit{XMM-Newton} observations of HD\,93250.}\\
\hline\hline
\centering
 Sr. 	& Obs. ID & Detector  & Obs. Date  & Start time  & Duration\tablefootmark{b}   & Livetime\tablefootmark{c}   &Source\tablefootmark{d}     & Offset\tablefootmark{e}  & Observing\tablefootmark{f}      \\
 No.    &         & (filter\tablefootmark{a})  &            & (UT)        &  (sec)    & (sec)      & counts    & (\arcmin) & mode 	\\	
\hline
\endfirsthead
\caption{continued.}\\
\hline\hline
Sr. 	& Obs. ID & Detector  & Obs. Date  & Start time  & Duration\tablefootmark{b} & Livetime\tablefootmark{c}   &Source\tablefootmark{d}     & Offset\tablefootmark{e}  & Observing\tablefootmark{f}    \\
 No.    &         & (filter\tablefootmark{a})  &            & (UT)        &  (sec)    & (sec)      & counts    & (\arcmin) & mode 	\\	
\hline
\endhead
\hline
\endfoot
          1      & 0112580601    &     MOS1 (Th)        & 2000-07-26  & 04:59:20    & 36509 & 33110    &  5526$\pm$75   & 7.216  & PFW  \\
                 &               &     MOS2 (Th)       &             &             &       & 30020    &  4446$\pm$67   &      &  PPW2   \\      
          2      & 0112580701    &     MOS1 (Th)       & 2000-07-27  & 23:49:25    & 12425 & 10900    &  1731$\pm$42   & 7.216  & PFW   \\
                 &               &     MOS2 (Th)       &             &             &       & 7927     &  1176$\pm$35   &      &  PPW2    \\
          3      & 0109530101    &     MOS1 (Th)       & 2000-12-24  & 10:30:07    & 10002 & 8872     &  679$\pm$27    & 12.05  & PFW    \\ 
                 &               &     MOS2 (Th)       &             &             &       & 8876     &  603$\pm$25    &     &  PFW      \\
                 &               &     PN   (Th)       &             &             &       & 4330     &  726$\pm$27    &     & PFW     \\          
          4      & 0109530201    &     MOS1 (Th)       & 2000-12-28  & 06:11:57    & 9817  & 8671     &  609$\pm$25    & 12.05   & PFW  \\
                 &               &     MOS2 (Th)       &             &             &       & 8674     &  578$\pm$25    &   & PFW       \\
                 &               &     PN   (Th)       &             &             &       & 4207     &  870$\pm$30    &    & PFW      \\           
          5      & 0109530301    &     MOS1 (Th)       & 2000-12-31  & 16:14:59    & 9909  & 8770     &  627$\pm$25    & 12.05  & PFW  \\
                 &               &     MOS2 (Th)       &             &             &       & 8771     &  483$\pm$23    &    & PFW      \\
                 &               &     PN   (Th)       &             &             &       & 4276     &  764$\pm$28    &     & PFW       \\                                     
          6      & 0109530401    &     MOS1 (Th)       & 2001-02-03  & 20:40:27    & 12716 & 11490    & 1025$\pm$33    & 12.05  & PFW  \\
                 &               &     MOS2 (Th)       &             &             &       & 11500    &  965$\pm$32    &    & PFW      \\
                 &               &     PN   (Th)       &             &             &       & 6640     &  1675$\pm$42   &     & PFW     \\               
          7      & 0112560101    &     MOS1 (Th)       & 2001-06-25  & 06:51:26    & 37052 & 25690    &  2199$\pm$49   & 10.25  & PFW  \\
                 &               &     MOS2 (Th)       &             &             &       & 25210    &  1938$\pm$45   &    & PFW      \\  
                 &               &     PN   (Th)       &             &             &       & 21430    &  5349$\pm$76   &    & PFW      \\                                        
          8      & 0112560201    &     MOS1 (Th)       & 2001-06-28  & 07:22:56    & 40092 & 24130    &  1945$\pm$45   & 10.25 & PFW  \\
                 &               &     MOS2 (Th)       &             &             &       & 24130    &  1722$\pm$44   &   & PFW       \\  
                 &               &     PN   (Th)       &             &             &       & 19800    &  5064$\pm$75   &    & PFW      \\
          9      & 0112560301    &     MOS1 (Th)       & 2001-06-30  & 04:39:01    & 37714 & 36320    &  2557$\pm$52   & 10.25  & PFW  \\
                 &               &     MOS2 (Th)       &             &             &       & 36340    &  2472$\pm$52   &    & PFW      \\     
                 &               &     PN   (Th)       &             &             &       & 30160    &  7413$\pm$90   &    & PFW      \\
         10      & 0145740101    &     MOS1 (Th)       & 2003-01-25  & 12:57:37    & 7252  &  6896    &   703$\pm$27   &  7.546  & PFW  \\
                 &               &     MOS2 (Th)       &             &             &       &  6906    &   719$\pm$28   &   & PFW       \\              
         11      & 0145740201    &     MOS1 (Th)       & 2003-01-27  & 01:02:55    & 7250  &  6898    &   682$\pm$27   &  7.546 & PFW  \\
                 &               &     MOS2 (Th)       &             &             &       &  6905    &   738$\pm$28   &     & PFW     \\              
         12      & 0145740301    &     MOS1 (Th)       & 2003-01-27  & 20:36:23    & 7247  &  6848    &   755$\pm$28   &  7.546  & PFW  \\
                 &               &     MOS2 (Th)       &             &             &       &  6851    &   815$\pm$29   &     & PFW     \\              
         13      & 0145740401    &     MOS1 (Th)       & 2003-01-29  & 01:39:47    & 8751  &  8380    &   890$\pm$30   &  7.546  & PFW  \\  
                 &               &     MOS2 (Th)       &             &             &       &  8389    &   938$\pm$32   &    & PFW      \\              
         14      & 0145740501    &     MOS1 (Th)       & 2003-01-29  & 23:54:24    & 7249  &  6896    &   719$\pm$27   &  7.546  & PFW  \\ 
                 &               &     MOS2 (Th)       &             &             &       &  6903    &   785$\pm$29   &    & PFW      \\              
         15      & 0160160101    &     MOS1 (Th)       & 2003-06-08  & 13:29:39    & 38352 & 16370    &  1792$\pm$43   &  7.546  & PPW2 \\
                 &               &     MOS2 (Th)       &             &             &       & 15590    &  1520$\pm$40   &    & PFW      \\                        
         16      & 0160160901    &     MOS1 (Th)       & 2003-06-13  & 23:51:15    & 31655 & 31090    &  3311$\pm$59   &  7.546   & PPW2  \\
                 &               &     MOS2 (Th)       &             &             &       & 31110    &  2570$\pm$52   &    & PFW      \\                       
         17      & 0145780101    &     MOS1 (Th)       & 2003-07-22  & 01:50:45    & 8736  &  8363    &   961$\pm$31   &  7.546  & PPW2  \\
                 &               &     MOS2 (Med)      &             &             &       &  8384    &   957$\pm$32   &     & PFW     \\                                
         18      & 0160560101    &     MOS1 (Th)       & 2003-08-02  & 21:00:17    & 17952 & 15880    &  1855$\pm$44   &  7.546   & PPW2 \\
                 &               &     MOS2 (Med)      &             &             &       & 11750    &  1574$\pm$41   &    & PFW      \\                                
         19      & 0160560201    &     MOS1 (Th)       & 2003-08-09  & 01:43:21    & 12952 & 12040    &  1474$\pm$39   &  7.546   & PFW  \\
         20      & 0160560301    &     MOS1 (Th)       & 2003-08-18  & 15:22:43    & 19143 & 18520    &  2446$\pm$50   &  7.546    & PPW2 \\
                 &               &     MOS2 (Med)      &             &             &       & 18570    &  2789$\pm$54   &   & PFW       \\                                
         21      & 0311990101    &     MOS1 (Th)       & 2006-01-31  & 18:03:33    & 66949 & 35510    &  3527$\pm$61   &  7.216  & PPW2  \\
                 &               &     MOS2 (Th)       &             &             &       & 35590    &  3704$\pm$62   &    & PPW2      \\                                
                 &               &     PN   (Th)       &             &             &       & 24010    &  7856$\pm$91   &    & PFW      \\
         22      & 0560580101    &     MOS1 (Th)       & 2009-01-05  & 10:22:08    & 14916 & 13990    &  1701$\pm$42   &  7.534   & PPW2 \\
                 &               &     MOS2 (Th)       &             &             &       & 14040    &  1815$\pm$44   &   & PFW       \\                                         
         23      & 0560580201    &     MOS1 (Th)       & 2009-01-09  & 14:27:16    & 11910 & 11420    &  1301$\pm$37   &  7.534   & PPW2  \\
                 &               &     MOS2 (Th)       &             &             &       & 11450    &  1316$\pm$38   &   & PFW      \\                                         
         24      & 0560580301    &     MOS1 (Th)       & 2009-01-15  & 11:22:01    & 26917 & 26110    &  3167$\pm$57   &  7.534  & PPW2 \\
                 &               &     MOS2 (Th)       &             &             &       & 26170    &  3315$\pm$59   &   & PFW       \\                                         
         25      & 0560580401    &     MOS1 (Th)       & 2009-02-02  & 04:45:24    & 26917 & 23490    &  2816$\pm$53   &  7.534   & PPW3  \\ 
                 &               &     MOS2 (Th)       &             &             &       & 26210    &  3181$\pm$58   &   & PFW     \\
                 &               &     PN   (Th)       &             &             &       & 22780    &  8918$\pm$97   &   & PLW     \\
         26      & 0742850301    &     MOS2 (Med)      & 2014-06-06  & 19:13:05    & 14300 & 12810    &  1810$\pm$43   &  7.534    & PFW  \\ 
         27      & 0742850401    &     MOS1 (Th)       & 2014-07-28  & 15:32:43    & 35000 & 33340    &  3187$\pm$57   &  7.534  & PPW2   \\ 
                 &               &     MOS2 (Med)      &             &             &       & 33320    &  4035$\pm$65   &  & PFW      \\         
         28      & 0762910401    &     MOS1 (Th)       & 2015-07-16  & 01:18:44    & 13000 & 11530    &  1550$\pm$40   &  7.526    & PPW2  \\
                 &               &     MOS2 (Med)      &             &             &       & 11500    &  1498$\pm$40   &   & PFW     \\         
         29      & 0804950201    &     MOS1 (Med)      & 2017-06-04  & 20:47:06    & 33000 & 31200    &  3756$\pm$63   &  6.121   & PPW2   \\
                 &               &     MOS2 (Med)      &             &             &       & 31220    &  3799$\pm$63   &   & PPW2     \\         
         30      & 0804950301    &     MOS1 (Med)      & 2017-12-06  & 06:26:36    & 30000 & 20620    &  2977$\pm$56   &  6.121   & PPW2   \\
                 &               &     MOS2 (Med)      &             &             &       & 20140    &  3011$\pm$56   &   & PPW2    \\         
         31      & 0830191801    &     MOS2 (Med)      & 2018-08-22  & 03:14:17    & 33100 & 29300    &  4114$\pm$65   &  6.121    & PPW2  \\
         32      & 0845030201    &     MOS1 (Med)      & 2019-06-07  & 05:25:44    & 34300 & 32400    &  3813$\pm$63   &  6.121   & PFW   \\
                 &               &     MOS2 (Med)      &             &             &       & 32450    &  3518$\pm$61   &     & PFW   \\         
         33      & 0845030301    &     MOS1 (Med)      & 2019-12-07  & 17:23:47    & 28000 & 25790    &  3629$\pm$62   &  6.121    & PFW   \\
                 &               &     MOS2 (Med)      &             &             &       & 26240    &  3560$\pm$61   &    & PFW    \\
                 &               &     PN   (Med)      &             &             &       & 16000    &  7110$\pm$86   &    & PFW    \\            
\end{longtable}
\tablefoottext{a}{`Th' and `Med' stand for the thick and medium optical blocking filters, respectively.} \\
\tablefoottext{b}{Total duration of the observation.} \\
\tablefoottext{c}{LIVETIME is the livetime keyword value corrected for periods of dead time and high background (wherever located).} \\
\tablefoottext{d}{Background corrected net source counts have been estimated in 0.3-10.0 keV energy range.} \\
\tablefoottext{e}{Offset between HD\,93250 position at EPIC instruments and the telescope pointing.}  \\
\tablefoottext{f}{The observing modes of observations are (i) PFW: Prime Full Window, (ii) PLW: Prime Large Window (covering full field of view), (ii) PPW2: Prime Partial Window2 (small central window), (iii) PPW3: Prime Partial Window3 (large central window)}
}

\subsection{\textit{Nustar} observations of HD\,93250 }

\begin{figure}[h!]
\centering
	\includegraphics[width=0.8\columnwidth]{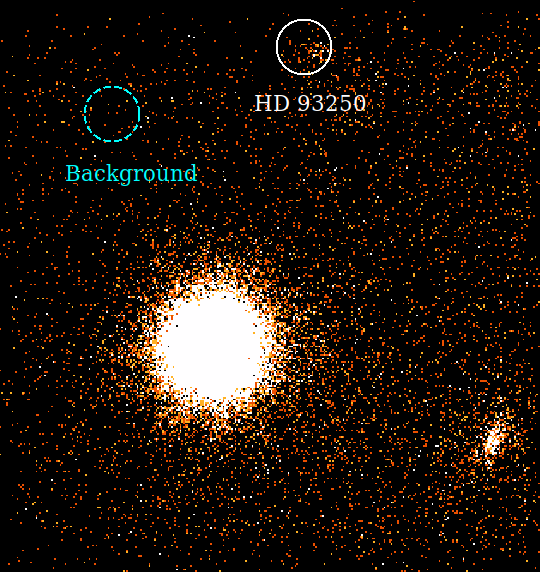}
	\caption{\textit{NuSTAR}$-$FPMA image of Eta-Carinae region in 3.0--78.0 keV energy range from observation ID 30002010012. The source and background regions have been highlighted with white and cyan colored circles, respectively, for the extraction of X-ray products. 
 \label{Nustar_image}}
\end{figure}

 \textit{NuSTAR} observed HD\,93250 on 8 occasions from 2014 July to 2020 August using both the focal plane modules (FPMA and FPMB). Since most of these observations were performed for Eta Carinae, our target source is present at large offsets from the detector aim-point as shown in the \textit{NuSTAR} log of observations in Table \ref{nustar_table}. Processing of \textit{NuSTAR} data was done using the software \textsc{nustardas v2.1.0} distributed by HEASARC within HEASoft 6.29. The task \textsc{nupipeline} (version 0.4.9) was used to create calibrated, cleaned, and screened event files with \textit{NuSTAR} \textsc{caldb} version 20231121. To extract the source and background counts, circular regions of 40 arcsec radius at a source and nearby surrounding source-free regions, respectively, on the same detector were selected as shown in Fig. \ref{Nustar_image}. Light curve and spectrum in 3–78 keV energy range and corresponding response files were extracted using \textsc{nuproducts} package within \textsc{nustardas} with time binning of 10 s. The source counts in different energy bands were estimated and have been mentioned in Table \ref{nustar_table}.

\begin{table*}
\caption{\label{nustar_table} Log of \textit{NuSTAR} observations of HD\,93250.}
\centering    
\setlength{\tabcolsep}{8.5pt}   
\renewcommand{\arraystretch}{1.5}
\begin{tabular}{c c c c c c c c c c}  
  \hline\hline  
 Sr. 	& Obs. ID & Orbital &Detector  & Obs. Date  & Start time  &  Livetime\tablefootmark{a}  & Offset\tablefootmark{b} &C1\tablefootmark{c}  & C2\tablefootmark{d}         \\
 \cline{9-10}
 No.    &     & phase    &   &            & (UT)        & (sec)    &  (\arcmin)     &    \multicolumn{2}{c}{(counts)}  	\\	
\hline
          1      &  30002040004   &  0.339  & FPMA        &   2014-07-28 & 10:31:07    & 61373    &  6.298   &  129$\pm$18      &     149$\pm$12     \\
                 &               &     & FPMB       &             &             &    61280   &     &      260$\pm$20   &    242$\pm$15    \\      
          2      & 30002010007    &   0.409  & FPMA       &   2014-08-11 & 05:36:07    & 31033    & 6.949    &    61$\pm$11   &    71$\pm$8     \\
                 &               &     & FPMB       &             &             &    31010   &      &       103$\pm$13   &    117$\pm$11        \\
          3      & 30002010008   &    0.414  & FPMA       &   2014-08-11 & 23:01:07    & 56943 &  6.337    &      129$\pm$18   &    231$\pm$16        \\ 
                 &               &     &  FPMB       &             &             &   56870    &      &      248$\pm$19   &    246$\pm$16       \\
          4      &  30002010010   &   0.454  &  FPMA       &  2014-08-19 & 16:41:07    & 54524  &  5.755    &      115$\pm$18   &    211$\pm$15   \\
                 &               &     &   FPMB       &             &             &  54480     &      &      177$\pm$18   &    213$\pm$15       \\   
          5      &  30002010012   &   0.647   &  FPMA       &  2014-09-26 & 00:41:07    & 81454  & 5.009     &      200$\pm$22   &    279$\pm$17 \\
                 &               &     &   FPMB       &             &             &  81320     &      &      306$\pm$24   &    282$\pm$17       \\
          6      &   30402001004  &  0.974    & FPMA       &   2018-08-20 & 00:36:09    &  33075  & 7.814    &      131$\pm$15   &    143$\pm$12   \\
                 &               &     &   FPMB       &             &             &  32890     &     &      136$\pm$16   &    122$\pm$11         \\            
          7      &  90501354002   &  0.418  & FPMA       &   2019-12-07 & 21:11:09  & 42457 &  7.230   &      166$\pm$18   &    182$\pm$13   \\
                 &               &     &   FPMB       &             &             &  42160     &     &      114$\pm$14   &    145$\pm$12       \\  
          8      &  30602030004   &   0.756  &  FPMA       & 2020-08-23 & 12:06:09    & 102221 &  5.443   &      312$\pm$28   &    408$\pm$20  \\
                 &               &     &   FPMB       &             &             &   101400     &     &      442$\pm$27   &    407$\pm$20        \\  
\hline
\end{tabular}
\tablefoot{(a) LIVETIME is the exposure time corrected for periods of dead time. \\
(b) Offset between HD\,93250 position at the detector plane and the telescope pointing. \\
(c) Background corrected net source counts in 3.0--10.0 keV energy range. \\
(d) Total counts in 10.0--78.0 keV at the position of HD\,93250 without background subtraction.\\
}
\end{table*}


\section{X-ray light curve analysis}\label{lc}
\begin{figure}
\centering
\subfigure{\includegraphics[width=0.98\columnwidth,trim={0.0cm 1.5cm 0.0cm 4.0cm}]{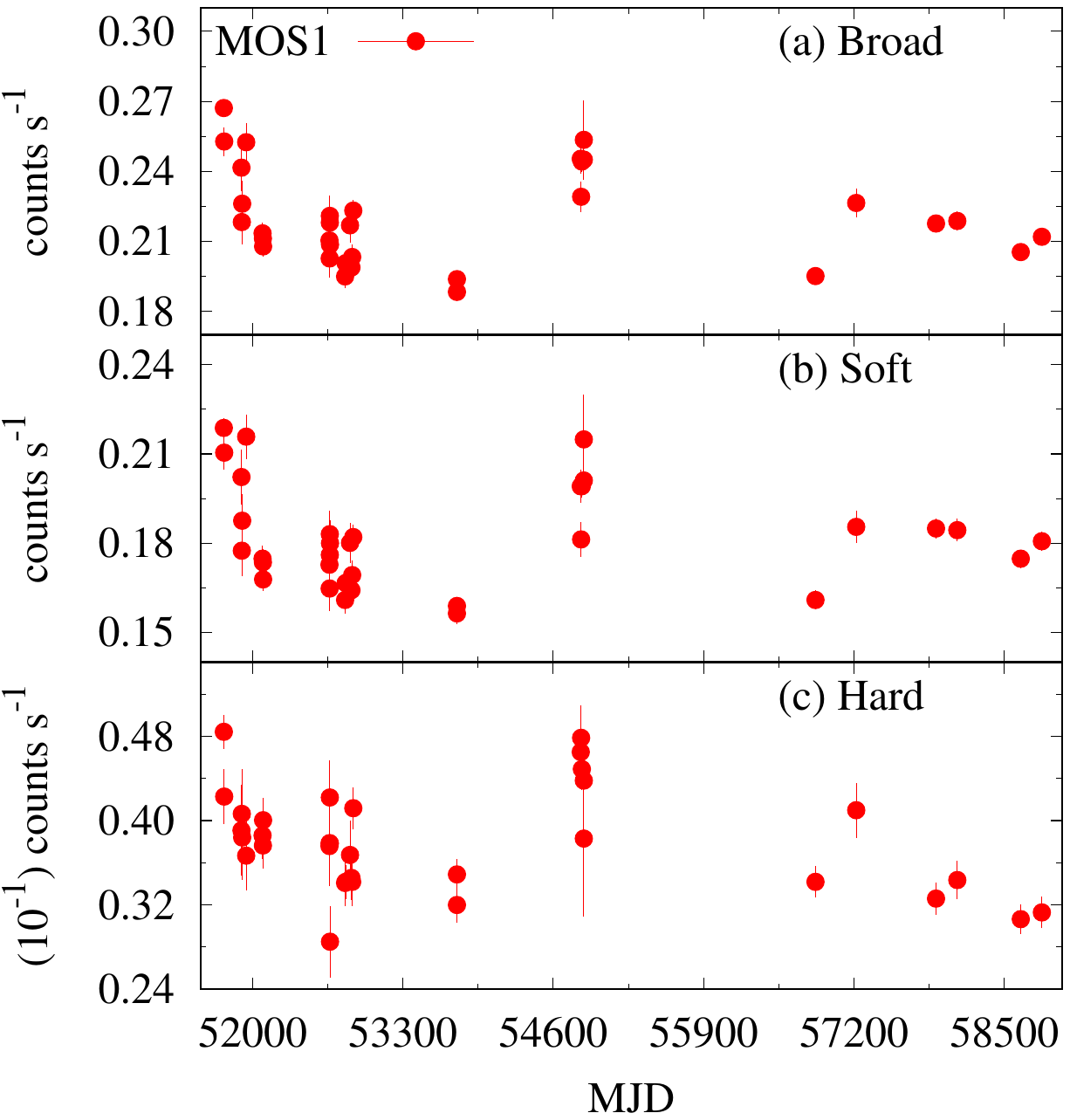}}
\caption{X-ray light curves of HD\,93250 as observed by \textit{XMM-Newton} using MOS1 in broad (0.3-10.0 keV), soft (0.3-2.0 keV), and hard (2.0-10.0 keV) energy bands. Each point corresponds to average count rate of an individual observation ID.
\label{fig:hd93250_lc}}
\end{figure}

The background-subtracted X-ray light curves as observed by MOS1 on-board \textit{XMM-Newton} are shown in Fig. \ref{fig:hd93250_lc}. These light curves were extracted in broad (0.3-10.0 keV), soft (0.3-2.0 keV), and hard (2.0-10.0 keV) energy bands. The X-ray variability is seen in all the energy band light curves with a comparatively lesser count rate in the hard band.   

\begin{figure}
\centering
  \includegraphics[width=0.98\columnwidth]{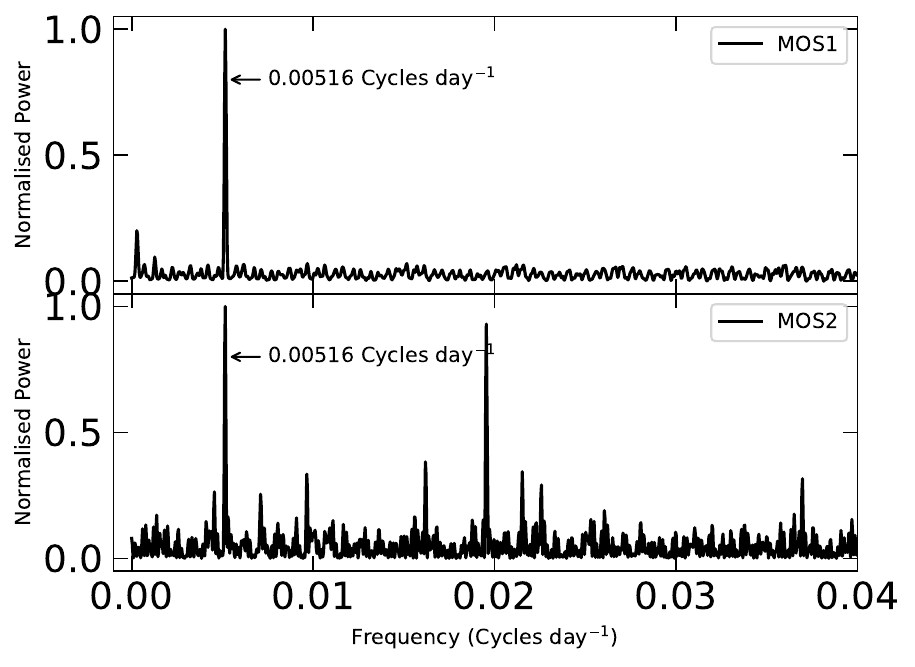}
\caption{CLEANed power spectra of HD\,93250 in the 0.3–10.0 keV energy range using timing data from \textit{XMM-Newton}--MOS1 and MOS2. The frequency of the peak with the highest power is also mentioned.
\label{periodogram}}
\end{figure}

We have performed a Fourier Transform (FT) of the MOS1 and MOS2 light curve in the 0.3-10.0 keV energy band to search for a periodic signal using the Lomb-Scargle periodogram \citep{1976Ap&SS..39..447L,1982ApJ...263..835S,1986ApJ...302..757H}. This is particularly effective in determining periodicity in the time series obtained over unequally spaced intervals of time and it showed maximum power at a frequency of (5.972$\pm$0.041)$\times$10$^{-8}$ s$^{-1}$ for both the light curves. However, several other peaks were also present in the Lomb-Scargle power spectra, which may be due to the aliasing. Therefore, we have also estimated the period using the CLEAN algorithm \cite{1987AJ.....93..968R} with a loop-gain of 0.1 and the number of iterations of 100. The resulting power spectra presented in the Fig. \ref{periodogram} show a dominant peak at a common frequency of 0.00516$\pm$0.00004 cycles per day (1$\sigma$ uncertainty) for MOS1 in the 0.3-10.0 keV energy band. Similar peak was noticed in the power spectra of MOS2. The consistent occurrence of the main peaks in both data sets at the same frequency lends strong support to its physical origin. However, most of the other features in the MOS2 power spectrum that are not displayed by MOS1 data are artifacts that can be disregarded in our analysis. The peak frequency corresponds to a period of 193.8$\pm$1.3 days and is consistent with the period derived by \citet{2017A&A...601A..34L}.  

\begin{figure*}
\centering
\subfigure[MOS1]{  \includegraphics[width=0.47\textwidth,trim={0.0cm 1.0cm 0.0cm 5.0cm}]{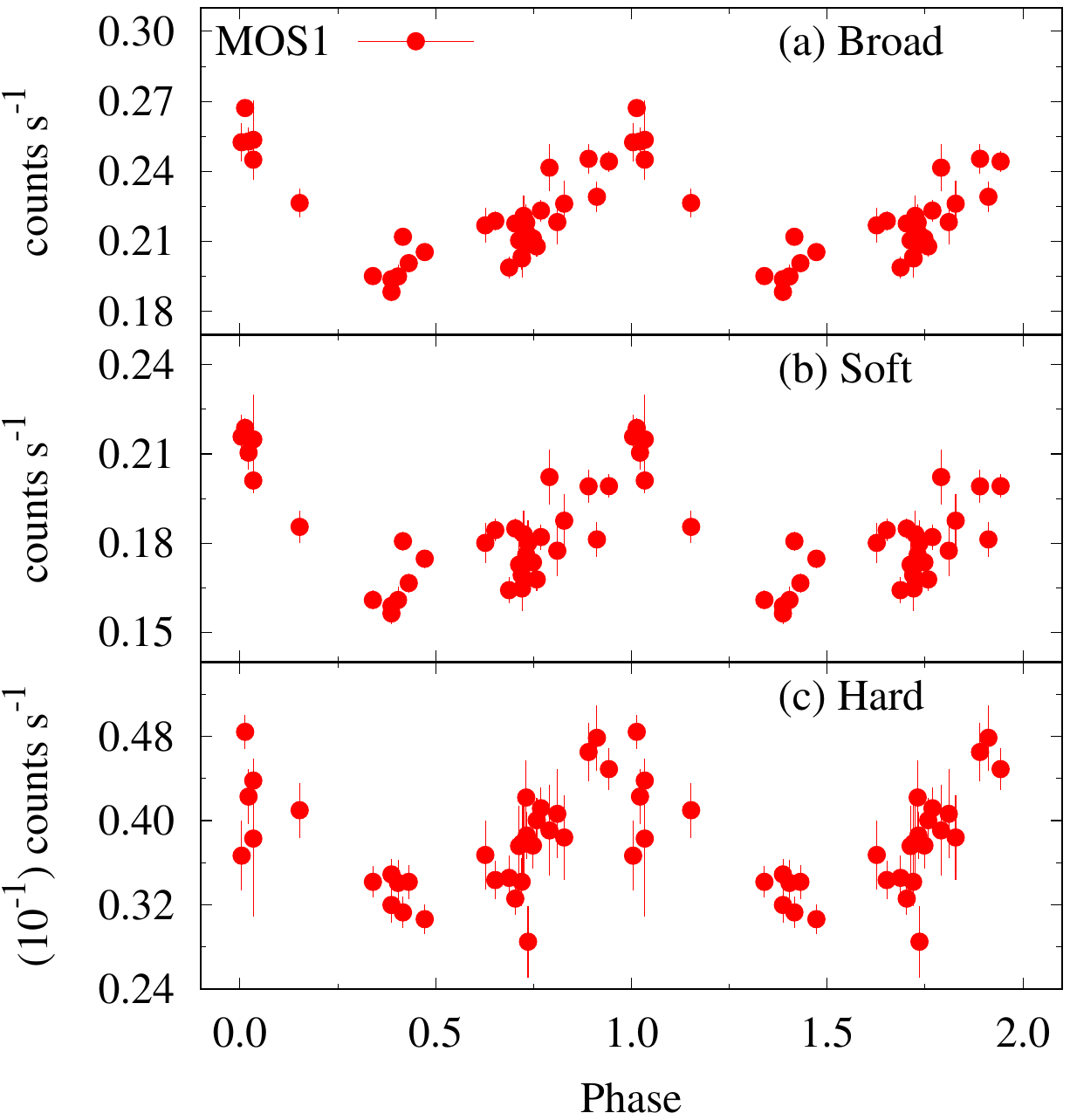}}
\subfigure[MOS2]{  \includegraphics[width=0.47\textwidth,trim={0.0cm 1.0cm 0.0cm 5.0cm}]{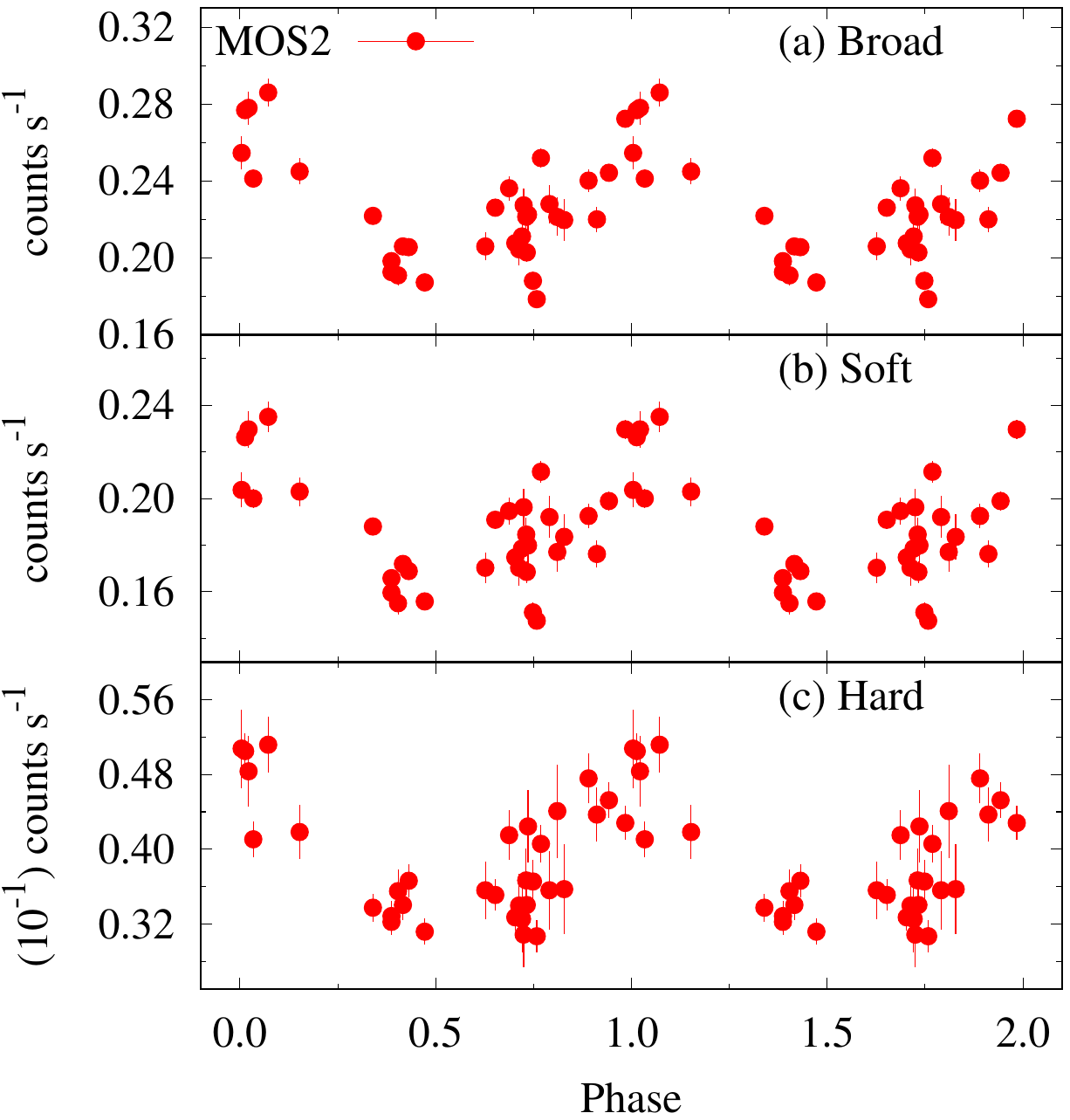}}
\caption{Folded X-ray light curves of HD\,93250 as observed by \textit{XMM-Newton} using (a) MOS1, and (b) MOS2 in broad, soft, and hard energy bands.
\label{fig:hd93250_flc}}
\end{figure*}

In order to ascertain the variation of count rate with the orbital phase ($\phi$), the light curves were folded using the ephemeris JD= 2454858.2$+$194.3E \citep{2017A&A...601A..34L}. The orbital phase has been estimated at the middle of the total observing time. Each point in the folded X-ray light curves, depicted in Fig. \ref{fig:hd93250_flc}, represents the average count rate of an observation ID observed by MOS1 and MOS2. Notably, the folded X-ray light curves of HD\,93250 exhibit phase-locked modulations, with the count rate peaking around orbital phase zero and gradually decreasing as the system approaches phase 0.5. This behavior is consistent across all three energy bands. The maximum-to-minimum count rate ratio in the broad, soft, and hard energy bands (with 1$\sigma$ uncertainty) is 1.38$\pm$0.03, 1.39$\pm$0.03, and 1.69$\pm$0.02 for MOS1, and 1.56$\pm$0.05, 1.55$\pm$0.06, and 1.66$\pm$0.02 for MOS2, respectively. It is worth noting that the difference in the detector effective area with thick and medium filters is only significant below ~0.5 keV for MOS detectors. Above 0.5 keV, filter thickness does not affect the count rates much. In addition, most of the measured X-ray emission from the target is measured above 0.5 keV, with only minor contribution down to 0.3 keV. Around 76\% of the MOS observations were performed with thick optical blocking filters (see Table \ref{tab:log}), with a rather good orbital sampling. The minute variations due to filter thickness should therefore not affect the trends visible in the phase folded X-ray light curves of HD\,93250.

\section{X-ray spectral analysis}\label{spec}

Given the nature of the system, we expect the X-ray emission to arise from both individual stellar winds \citep{1997A&A...322..878F,2013MNRAS.429.3379O}, and from the shocked gas in the colliding-wind region \citep{1992ApJ...386..265S,2010MNRAS.403.1657P}. In line with usual studies for massive star systems, the X-ray spectrum can be modeled using {\sc apec} optically thin thermal emission components \citep[e.g.][]{2005MNRAS.361..191S,2015MNRAS.451.1070D,2019MNRAS.487.2624A}. We adopted the reduced $\chi^2$ minimization approach as a goodness of fit criterion. We tested both two-temperature and three-temperature composite models, to account for the temperature distribution of the X-ray emitting plasma in the system. Overall, using only two emission components turned out to be insufficient to achieve reasonable modeling in the broad EPIC band. The values of reduced $\chi^2$ obtained for two-temperature components close to 2 dropped well below 1.4 for the three-temperature composite model. We thus focused on the three-temperature model. We also used two photoelectric absorption components ({\sc phabs}) to account respectively for the interstellar absorption and the local absorption by the stellar wind material. We determined the interstellar column using the Hydrogen Column Density Calculator \footnote{\url https://heasarc.gsfc.nasa.gov/cgi-bin/Tools/w3nh/w3nh.pl}, and we fixed the value to $N_{H}^{ISM}=0.36 \times 10^{22}$ cm$^{-2}$. The local column was left as a free parameter throughout all our fitting attempts. Our composite model was finally \textsc{phabs(ism)*phabs(local)*(apec+apec+apec)}. Our spectral modeling assumed default solar abundances in XSPEC \citep{1989GeCoA..53..197A}. We note that we also cross-checked with other abundance lists \citep{Asplund,Lodders}. However, this led to poorer constraints on the local absorbing column (mainly upper limits), preventing us from extracting valuable information on the varying column as a function of the orbital phase. This may be explained by the higher metallicity of recent abundance lists, leading the absorption component meant to account for ISM absorption to overestimate its actual contribution. For the sake of the physical consistency of our spectral analysis, we focused on the results allowing for a non-zero absorption by the stellar wind material. We stress that changing overall abundances should be taken with caution. More recent abundances arising from inner solar system measurements should by no means be viewed as improvement in the assumed abundances for a source located 2.3\,kpc away.

We froze some parameters to constrain the exploration of the parameter space and achieve a consistent set of solutions for the full-time series. 
We note that similar results were obtained for MOS1 and MOS2 spectra, and we focused on the simultaneous fitting of both MOS spectra of each epoch.

All emission model temperatures were frozen ($k$T$_{1}$ = 0.26 keV, $k$T$_{2}$ = 1.0 keV, and $k$T$_{3}$ = 3.3 keV), keeping free the local absorption column and the three normalization parameters. Following this approach, we obtained a consistent modeling series for the 33 epochs, including cases where the number of counts were too low to obtain adequate results in our first unconstrained attempts. All our results are presented in Table\,\ref{spec_par}. Reduced $\chi^2$ values range between 0.8 and 1.3. This is not significantly different from values obtained without any constrain on plasma temperatures, but the epoch-to-epoch consistency has been highly enhanced.

\begin{figure}
\centering
  \includegraphics[width=1.0\columnwidth,trim={0.0cm 0.0cm 4.0cm 1.0cm}]{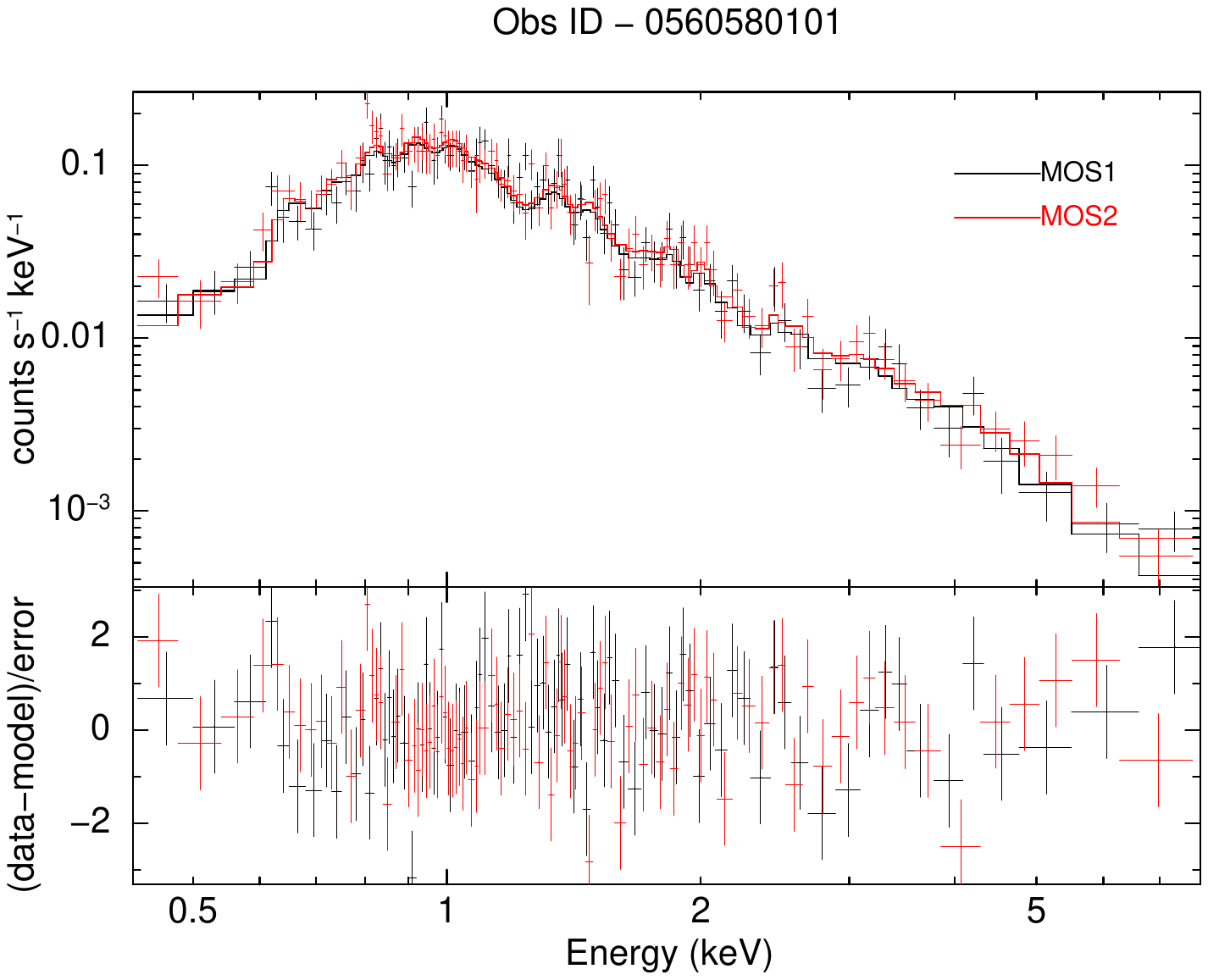}
\caption{MOS1 and MOS2 spectra of HD\,93250 jointly fitted with the three temperature thermal plasma emission model from the observation ID 0560580101. The details of the spectral fitting are provided in Table \ref{spec_par}.
\label{fig:hd93250_spectra}}
\end{figure}

\begin{table*}
\caption{\label{spec_par}Best fit parameters obtained from spectral fitting of HD\,93250 as observed from \textit{XMM-Newton}-EPIC.}
\centering    
\setlength{\tabcolsep}{1.2pt}   
\renewcommand{\arraystretch}{1.5}
\begin{tabular}{c c c c c c c c c c c c c}   
  \hline\hline
Obs. ID  & $\phi$    & $norm_{1}$        & $norm_{2}$        & $norm_{3}$  & N$_{H}^{local}$  &  $F^{obs}_{B}$ & $F^{obs}_{S}$ & $F^{obs}_{H}$ & $F^{ism}_{B}$ & $F^{ism}_{S}$ & $F^{ism}_{H}$ & $\chi^{2}_{\nu} (dof)$ \\ 
 \cline{4-5}                  \cline{7-12}                       
  &         & ($10^{-3}$ cm$^{-5}$)   & \multicolumn{2}{c}{($10^{-4}$ cm$^{-5}$)} & ($10^{22}$ cm$^{-2}$)   &   \multicolumn{6}{c}{($10^{-12}$ erg cm$^{-2}$ s$^{-1}$)} &    \\
  \hline
 0112580601 &0.013 &$2.91^{+0.74}_{-0.65}$ &  $6.23^{+0.39}_{-0.40}$  & $10.09^{+ 0.31}_{-0.31}$ &$0.16^{+0.04}_{-0.04}$  &$2.11^{+0.02}_{-0.02}$ &$ 1.18^{+0.01}_{-0.01}$&$0.93^{+0.01}_{-0.01}$ &$4.79^{+0.05}_{-0.05}$&$3.82^{+0.04}_{-0.04}$ &  $0.97^{+0.01}_{-0.01}$ &1.25 (368)\\
0112580701 &0.022 &$2.13^{+1.25}_{-1.19}$ &  $6.91^{+1.00}_{-1.54}$  & $9.33^{+ 0.68}_{-0.64}$ &$0.15^{+0.08}_{-0.13}$  &$1.99^{+0.04}_{-0.04}$ &$ 1.11^{+0.02}_{-0.02}$&$0.88^{+0.02}_{-0.02}$ &$4.38^{+0.09}_{-0.09}$&$3.46^{+0.07}_{-0.07}$ &  $0.92^{+0.02}_{-0.02}$ &1.24 (156)\\
0109530101 &0.791 &$5.58^{+2.51}_{-2.08}$ &  $4.55^{+0.99}_{-1.14}$  & $7.62^{+ 0.83}_{-0.83}$ &$0.29^{+0.08}_{-0.09}$  &$1.69^{+0.05}_{-0.05}$ &$ 1.01^{+0.03}_{-0.03}$&$0.69^{+0.02}_{-0.02}$ &$3.91^{+0.12}_{-0.12}$&$3.19^{+0.09}_{-0.09}$ &  $0.72^{+0.02}_{-0.02}$ &1.21 ( 76)\\
0109530201 &0.810 &$6.15^{+2.36}_{-1.97}$ &  $4.32^{+1.12}_{-1.26}$  & $8.93^{+ 0.89}_{-0.89}$ &$0.35^{+0.07}_{-0.08}$  &$1.70^{+0.05}_{-0.05}$ &$ 0.92^{+0.03}_{-0.03}$&$0.78^{+0.02}_{-0.02}$ &$3.51^{+0.11}_{-0.11}$&$2.69^{+0.08}_{-0.08}$ &  $0.82^{+0.03}_{-0.03}$ &0.93 ( 71)\\
0109530301 &0.828 &$7.07^{+2.64}_{-2.29}$ &  $4.86^{+1.29}_{-1.53}$  & $7.98^{+ 0.97}_{-0.94}$ &$0.38^{+0.07}_{-0.08}$  &$1.65^{+0.05}_{-0.05}$ &$ 0.93^{+0.03}_{-0.03}$&$0.72^{+0.02}_{-0.02}$ &$3.44^{+0.11}_{-0.11}$&$2.69^{+0.09}_{-0.09}$ &  $0.75^{+0.02}_{-0.02}$ &0.80 ( 67)\\
0109530401 &0.004 &$5.54^{+1.99}_{-1.69}$ &  $5.53^{+0.85}_{-0.93}$  & $9.16^{+ 0.76}_{-0.75}$ &$0.29^{+0.06}_{-0.07}$  &$1.91^{+0.05}_{-0.05}$ &$ 1.08^{+0.03}_{-0.03}$&$0.83^{+0.02}_{-0.02}$ &$4.19^{+0.10}_{-0.10}$&$3.32^{+0.08}_{-0.08}$ &  $0.87^{+0.02}_{-0.02}$ &1.14 (112)\\
0112560101 &0.732 &$5.17^{+1.07}_{-0.97}$ &  $4.90^{+0.52}_{-0.54}$  & $7.22^{+ 0.41}_{-0.41}$ &$0.32^{+0.04}_{-0.04}$  &$1.57^{+0.03}_{-0.03}$ &$ 0.91^{+0.02}_{-0.02}$&$0.66^{+0.01}_{-0.01}$ &$3.43^{+0.06}_{-0.06}$&$2.73^{+0.05}_{-0.04}$ &  $0.69^{+0.01}_{-0.01}$ &1.05 (214)\\
0112560201 &0.748 &$3.45^{+0.91}_{-0.79}$ &  $4.15^{+0.44}_{-0.43}$  & $7.62^{+ 0.41}_{-0.41}$ &$0.23^{+0.05}_{-0.05}$  &$1.55^{+0.03}_{-0.03}$ &$ 0.87^{+0.02}_{-0.02}$&$0.68^{+0.01}_{-0.01}$ &$3.47^{+0.06}_{-0.06}$&$2.76^{+0.05}_{-0.05}$ &  $0.72^{+0.01}_{-0.01}$ &1.00 (196)\\
0112560301 &0.758 &$2.49^{+0.81}_{-0.67}$ &  $4.37^{+0.37}_{-0.36}$  & $6.87^{+ 0.32}_{-0.33}$ &$0.19^{+0.05}_{-0.06}$  &$1.46^{+0.02}_{-0.02}$ &$ 0.83^{+0.01}_{-0.01}$&$0.63^{+0.01}_{-0.01}$ &$3.33^{+0.05}_{-0.05}$&$2.67^{+0.04}_{-0.04}$ &  $0.66^{+0.01}_{-0.01}$ &1.23 (246)\\
0145740101 &0.713 &$5.09^{+1.95}_{-1.63}$ &  $4.28^{+0.90}_{-1.03}$  & $7.96^{+ 0.76}_{-0.75}$ &$0.32^{+0.07}_{-0.08}$  &$1.59^{+0.05}_{-0.05}$ &$ 0.88^{+0.02}_{-0.03}$&$0.71^{+0.02}_{-0.02}$ &$3.39^{+0.10}_{-0.10}$&$2.65^{+0.07}_{-0.08}$ &  $0.74^{+0.02}_{-0.02}$ &0.80 ( 84)\\
0145740201 &0.720 &$3.84^{+2.05}_{-1.59}$ &  $4.26^{+0.76}_{-0.84}$  & $7.56^{+ 0.69}_{-0.69}$ &$0.25^{+0.09}_{-0.10}$  &$1.57^{+0.04}_{-0.04}$ &$ 0.89^{+0.03}_{-0.03}$&$0.68^{+0.02}_{-0.02}$ &$3.53^{+0.10}_{-0.10}$&$2.82^{+0.08}_{-0.08}$ &  $0.71^{+0.02}_{-0.02}$ &0.96 ( 85)\\
0145740301 &0.725 &$5.88^{+2.49}_{-2.09}$ &  $7.19^{+1.01}_{-1.10}$  & $6.73^{+ 0.74}_{-0.75}$ &$0.36^{+0.07}_{-0.09}$  &$1.63^{+0.04}_{-0.04}$ &$ 0.97^{+0.03}_{-0.03}$&$0.66^{+0.02}_{-0.02}$ &$3.47^{+0.09}_{-0.09}$&$2.78^{+0.07}_{-0.08}$ &  $0.69^{+0.02}_{-0.02}$ &1.06 ( 91)\\
0145740401 &0.731 &$7.31^{+2.59}_{-2.31}$ &  $4.59^{+1.03}_{-1.23}$  & $8.47^{+ 0.75}_{-0.73}$ &$0.39^{+0.07}_{-0.08}$  &$1.67^{+0.04}_{-0.04}$ &$ 0.92^{+0.02}_{-0.02}$&$0.75^{+0.02}_{-0.02}$ &$3.41^{+0.08}_{-0.09}$&$2.62^{+0.07}_{-0.07}$ &  $0.78^{+0.02}_{-0.02}$ &1.09 (105)\\
0145740501 &0.736 &$3.84^{+1.80}_{-1.45}$ &  $6.21^{+0.85}_{-0.79}$  & $6.92^{+ 0.68}_{-0.69}$ &$0.28^{+0.08}_{-0.09}$  &$1.58^{+0.04}_{-0.04}$ &$ 0.92^{+0.03}_{-0.03}$&$0.66^{+0.02}_{-0.02}$ &$3.42^{+0.09}_{-0.09}$&$2.72^{+0.08}_{-0.08}$ &  $0.69^{+0.02}_{-0.02}$ &1.33 ( 87)\\
0160160101 &0.403 &$1.69^{+0.86}_{-0.62}$ &  $3.53^{+0.43}_{-0.43}$  & $7.22^{+ 0.36}_{-0.37}$ &$0.09^{+0.07}_{-0.08}$  &$1.49^{+0.03}_{-0.03}$ &$ 0.85^{+0.02}_{-0.02}$&$0.65^{+0.01}_{-0.01}$ &$3.74^{+0.07}_{-0.07}$&$3.06^{+0.06}_{-0.06}$ &  $0.68^{+0.01}_{-0.01}$ &1.04 (175)\\
0160160901 &0.431 &$4.07^{+0.77}_{-0.69}$ &  $4.07^{+0.35}_{-0.36}$  & $7.10^{+ 0.31}_{-0.31}$ &$0.25^{+0.04}_{-0.04}$  &$1.53^{+0.02}_{-0.02}$ &$ 0.89^{+0.01}_{-0.01}$&$0.64^{+0.01}_{-0.01}$ &$3.54^{+0.05}_{-0.05}$&$2.87^{+0.04}_{-0.04}$ &  $0.67^{+0.01}_{-0.01}$ &1.12 (273)\\
0145780101 &0.626 &$3.95^{+1.57}_{-1.36}$ &  $4.57^{+0.65}_{-0.64}$  & $6.83^{+ 0.55}_{-0.56}$ &$0.28^{+0.07}_{-0.08}$  &$1.46^{+0.04}_{-0.04}$ &$ 0.83^{+0.02}_{-0.02}$&$0.63^{+0.02}_{-0.02}$ &$3.19^{+0.08}_{-0.08}$&$2.53^{+0.06}_{-0.06}$ &  $0.66^{+0.02}_{-0.02}$ &1.27 (113)\\
0160560101 &0.687 &$4.46^{+1.25}_{-1.10}$ &  $4.46^{+0.51}_{-0.54}$  & $7.34^{+ 0.44}_{-0.44}$ &$0.29^{+0.05}_{-0.06}$  &$1.54^{+0.03}_{-0.03}$ &$ 0.87^{+0.02}_{-0.02}$&$0.66^{+0.01}_{-0.01}$ &$3.38^{+0.06}_{-0.06}$&$2.68^{+0.05}_{-0.05}$ &  $0.69^{+0.01}_{-0.01}$ &1.19 (181)\\
0160560201 &0.719 &$5.68^{+2.69}_{-2.51}$ &  $3.95^{+0.89}_{-1.16}$  & $7.47^{+ 0.68}_{-0.66}$ &$0.33^{+0.09}_{-0.12}$  &$1.54^{+0.04}_{-0.04}$ &$ 0.88^{+0.02}_{-0.02}$&$0.66^{+0.02}_{-0.02}$ &$3.36^{+0.09}_{-0.09}$&$2.67^{+0.07}_{-0.07}$ &  $0.69^{+0.02}_{-0.02}$ &0.96 ( 79)\\
0160560301 &0.768 &$3.44^{+1.01}_{-0.89}$ &  $5.04^{+0.39}_{-0.37}$  & $8.20^{+ 0.36}_{-0.36}$ &$0.23^{+0.05}_{-0.06}$  &$1.70^{+0.02}_{-0.02}$ &$ 0.96^{+0.01}_{-0.01}$&$0.75^{+0.01}_{-0.01}$ &$3.79^{+0.05}_{-0.05}$&$3.00^{+0.04}_{-0.04}$ &  $0.78^{+0.01}_{-0.01}$ &1.18 (250)\\
0311990101 &0.387 &$2.99^{+0.79}_{-0.75}$ &  $4.54^{+0.30}_{-0.29}$  & $6.76^{+ 0.27}_{-0.27}$ &$0.21^{+0.05}_{-0.05}$  &$1.48^{+0.02}_{-0.02}$ &$ 0.86^{+0.01}_{-0.01}$&$0.62^{+0.01}_{-0.01}$ &$3.41^{+0.041}_{-0.04}$&$2.76^{+0.03}_{-0.03}$ &  $0.65^{+0.01}_{-0.01}$ &1.32 (312)\\
0560580101 &0.890 &$2.59^{+1.04}_{-0.82}$ &  $4.82^{+0.52}_{-0.49}$  & $9.43^{+ 0.46}_{-0.47}$ &$0.15^{+0.06}_{-0.07}$  &$1.89^{+0.03}_{-0.03}$ &$ 1.05^{+0.02}_{-0.02}$&$0.85^{+0.02}_{-0.02}$ &$4.37^{+0.08}_{-0.08}$&$3.48^{+0.06}_{-0.06}$ &  $0.89^{+0.02}_{-0.02}$ &1.01 (187)\\
0560580201 &0.912 &$2.61^{+1.47}_{-1.20}$ &  $4.81^{+0.65}_{-0.72}$  & $9.55^{+ 0.53}_{-0.54}$ &$0.19^{+0.09}_{-0.11}$  &$1.79^{+0.04}_{-0.04}$ &$ 0.94^{+0.02}_{-0.02}$&$0.85^{+0.02}_{-0.02}$ &$3.80^{+0.08}_{-0.08}$&$2.91^{+0.06}_{-0.06}$ &  $0.89^{+0.02}_{-0.02}$ &1.06 (146)\\
0560580301 &0.942 &$2.85^{+0.98}_{-0.91}$ &  $5.53^{+0.47}_{-0.54}$  & $9.34^{+ 0.37}_{-0.37}$ &$0.18^{+0.06}_{-0.07}$  &$1.90^{+0.02}_{-0.02}$ &$ 1.05^{+0.01}_{-0.01}$&$0.85^{+0.01}_{-0.01}$ &$4.26^{+0.06}_{-0.06}$&$3.37^{+0.04}_{-0.04}$ &  $0.89^{+0.01}_{-0.01}$ &1.13 (297)\\
0560580401 &0.034 &$2.26^{+0.81}_{-0.72}$ &  $5.76^{+0.53}_{-0.61}$  & $8.81^{+ 0.38}_{-0.38}$ &$0.14^{+0.06}_{-0.07}$  &$1.88^{+0.03}_{-0.03}$ &$ 1.06^{+0.01}_{-0.01}$&$0.82^{+0.01}_{-0.01}$ &$4.32^{+0.06}_{-0.06}$&$3.47^{+0.05}_{-0.05}$ &  $0.86^{+0.01}_{-0.01}$ &1.04 (282)\\
0742850301 &0.072 &$4.17^{+1.99}_{-1.60}$ &  $5.49^{+0.84}_{-0.77}$  & $10.37^{+ 0.74}_{-0.75}$ &$0.24^{+0.08}_{-0.09}$ &$2.02^{+0.05}_{-0.05}$ &$ 1.09^{+0.03}_{-0.03}$&$0.93^{+0.02}_{-0.02}$ &$4.34^{+0.11}_{-0.11}$&$3.37^{+0.08}_{-0.08}$ &  $0.97^{+0.02}_{-0.02}$ &0.78 ( 95)\\
0742850401 &0.339 &$2.25^{+0.68}_{-0.61}$ &  $3.95^{+0.28}_{-0.29}$  & $6.91^{+ 0.26}_{-0.26}$ &$0.14^{+0.05}_{-0.06}$  &$1.49^{+0.02}_{-0.02}$ &$ 0.87^{+0.01}_{-0.01}$&$0.63^{+0.01}_{-0.01}$ &$3.66^{+0.04}_{-0.05}$&$2.99^{+0.04}_{-0.04}$ &  $0.66^{+0.01}_{-0.01}$ &1.17 (307)\\
0762910401 &0.152 &$9.83^{+0.50}_{-0.06}$ &  $3.33^{+0.46}_{-0.27}$  & $9.26^{+ 0.39}_{-0.42}$ &$0.07\dagger$        &$1.78^{+0.03}_{-0.03}$ &$ 0.96^{+0.02}_{-0.02}$&$0.82^{+0.02}_{-0.02}$ &$4.70^{+0.09}_{-0.09}$&$3.84^{+0.07}_{-0.07}$ &  $0.86^{+0.02}_{-0.02}$ &1.15 (166)\\
0804950201 &0.703 &$2.39^{+0.62}_{-0.58}$ &  $4.58^{+0.31}_{-0.32}$  & $6.32^{+ 0.26}_{-0.26}$ &$0.18^{+0.04}_{-0.05}$  &$1.42^{+0.02}_{-0.02}$ &$ 0.83^{+0.01}_{-0.01}$&$0.59^{+0.01}_{-0.01}$ &$3.29^{+0.04}_{-0.04}$&$2.67^{+0.03}_{-0.03}$ &  $0.62^{+0.01}_{-0.01}$ &1.19 (311)\\
0804950301 &0.652 &$2.59^{+1.02}_{-1.09}$ &  $4.33^{+0.37}_{-0.51}$  & $7.10^{+ 0.33}_{-0.32}$ &$0.19^{+0.06}_{-0.10}$  &$1.49^{+0.02}_{-0.02}$ &$ 0.85^{+0.01}_{-0.01}$&$0.65^{+0.01}_{-0.01}$ &$3.43^{+0.05}_{-0.05}$&$2.75^{+0.04}_{-0.04}$ &  $0.68^{+0.01}_{-0.01}$ &1.46 (271)\\
0830191801 &0.984 &$1.74^{+1.11}_{-0.73}$ &  $4.84^{+0.71}_{-0.83}$  & $9.09^{+ 0.47}_{-0.46}$ &$0.17\dagger$        &$1.87^{+0.03}_{-0.03}$ &$ 1.05^{+0.02}_{-0.02}$&$0.83^{+0.01}_{-0.01}$ &$4.53^{+0.07}_{-0.07}$&$3.66^{+0.06}_{-0.06}$ & $0.87^{+0.01}_{-0.01}$ &1.49 (166)\\
0845030201 &0.472 &$1.23^{+0.53}_{-0.39}$ &  $3.19^{+0.30}_{-0.35}$  & $6.13^{+ 0.22}_{-0.22}$ &$0.12\dagger$        &$1.32^{+0.02}_{-0.02}$ &$ 0.76^{+0.01}_{-0.01}$&$0.56^{+0.01}_{-0.01}$ &$3.43^{+0.04}_{-0.04}$&$2.85^{+0.03}_{-0.04}$ &  $0.58^{+0.01}_{-0.01}$ &1.44 (309)\\
0845030301 &0.416 &$0.91^{+0.25}_{-0.03}$ &  $3.15^{+0.24}_{-0.15}$  & $6.72^{+ 0.16}_{-0.22}$ &$0.04\dagger$        &$1.43^{+0.02}_{-0.02}$ &$ 0.82^{+0.01}_{-0.01}$&$0.61^{+0.01}_{-0.01}$ &$4.04^{+0.05}_{-0.05}$&$3.41^{+0.04}_{-0.04}$ &  $0.64^{+0.01}_{-0.01}$ &1.13 (315)\\
\hline
\end{tabular}
\tablefoot{Fit parameters are derived from joint spectral fitting of \textit{XMM-Newton}$-$MOS1 and MOS2 spectra of HD\,93250 using model \textsc{phabs(ism)*phabs(local)*(apec+apec+apec)} with fixed values of $N_{H}^{ISM}=0.36 \times 10^{22}$ $cm^{-2}$ and $k$T$_{1}$ = 0.26 keV, $k$T$_{2}$ = 1.0 keV, and $k$T$_{3}$ = 3.3 keV. $norm_{1}$, $norm_{2}$, and $norm_{3}$ are the normalization parameters for three temperature components whereas $N_{H}^{local}$ is the equivalent local H-column density. $F_{B}^{obs}$,  $F_{S}^{obs}$, and $F_{H}^{obs}$ are the observed and $F_{B}^{ism}$,  $F_{S}^{ism}$, and $F_{H}^{ism}$ are the ISM corrected X-ray fluxes of HD\,93250 in broad, soft, and hard energy bands, respectively. $\chi_{\nu}^{2}$ is the reduced $\chi^2$  and  \textit{dof} is degrees of freedom.  Errors quoted on different parameters refer to 90\% confidence level.  \\ $\dagger$ Mentioned values correspond to the upper limit on the specified parameter.}
\end{table*}

\begin{figure}
\centering
\includegraphics[width=0.99\columnwidth,trim={0.0cm 1.0cm 0.0cm 0.0cm}]{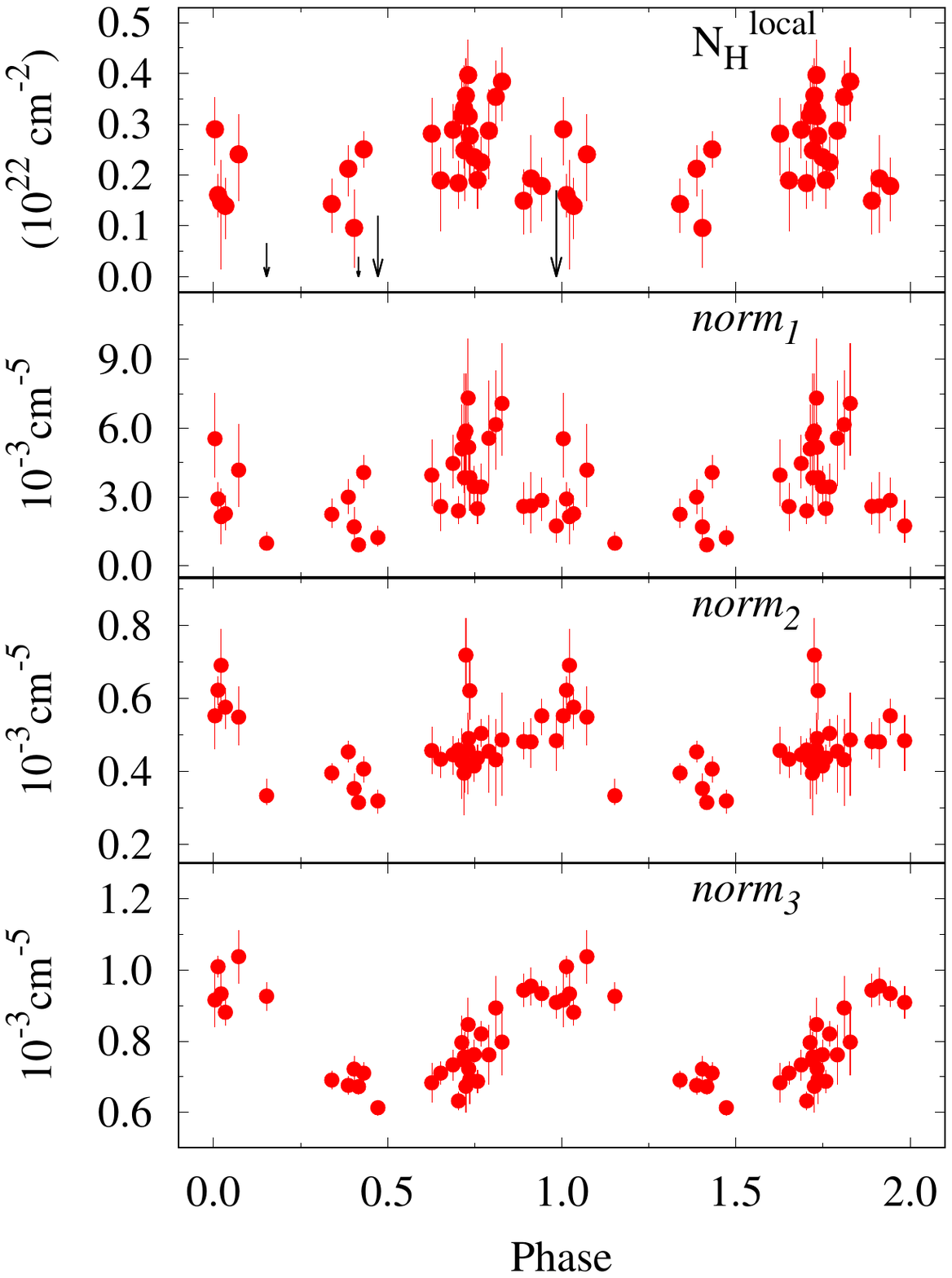}
\caption{Variation of the local equivalent H-column density (N$_{H}^{local}$) and normalization constants ($norm_{1}$, $norm_{2}$ and $norm_{3}$) corresponding to three thermal plasma emission components with the orbital phase of HD\,93250. The vertical arrows in the top most panel display the upper limit values of N$_{H}^{local}$ at specific orbital phases obtained after X-ray spectral fitting (see Table \ref{spec_par})}.
\label{fig:hd93250_spec_par}
\end{figure}

\begin{figure*}
\centering  
\subfigure[Observed flux]{\includegraphics[width=0.97\columnwidth,trim={0.4cm 1.5cm 0.0cm 5.0cm}]{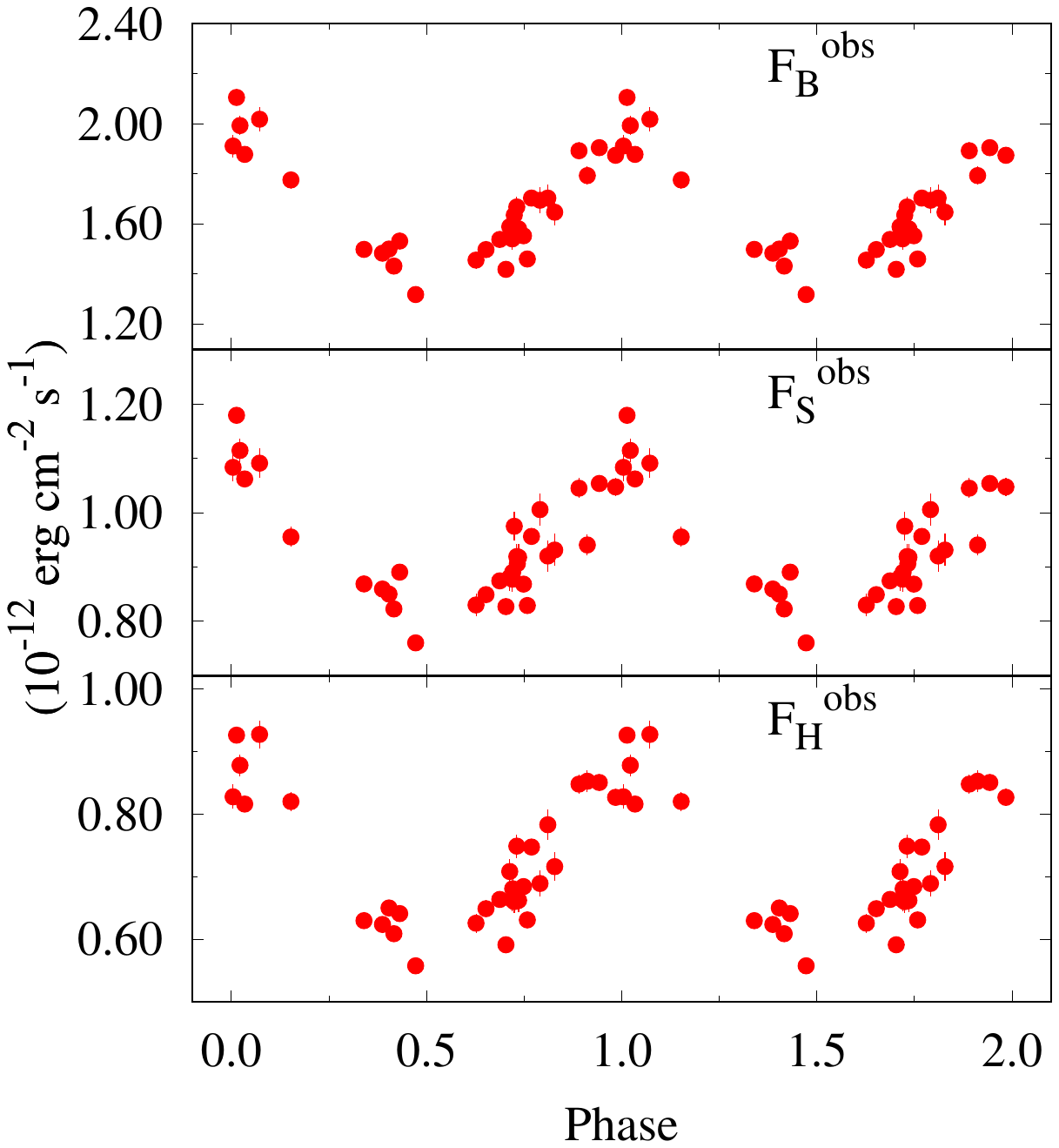}}
\subfigure[ISM corrected flux]{\includegraphics[width=0.97\columnwidth,trim={0.4cm 1.5cm 0.0cm 5.0cm}]{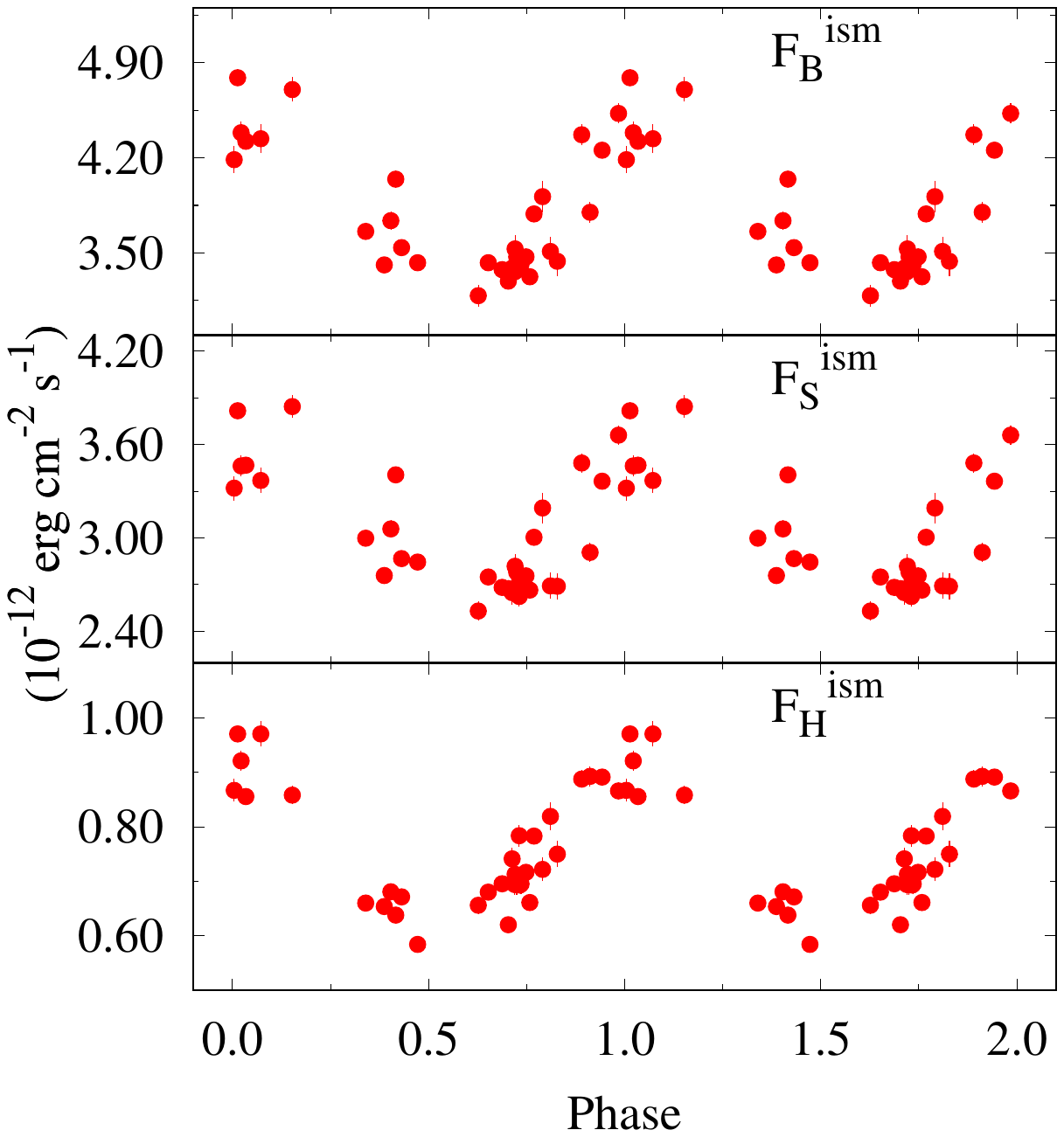}}
\caption{Variation of the (a) observed ($F^{obs}$) and (b) ism-corrected ($F^{ism}$) X-ray flux in broad, soft, and hard energy bands from HD\,93250 obtained after X-ray spectral fitting as a function of the binary orbital phase (see Table \ref{spec_par}).
\label{fig:flux_phase}}
\end{figure*}

\begin{figure}
\centering
  \includegraphics[width=1.0\columnwidth,trim={3.2cm 1.5cm 3.8cm 1.2cm}]{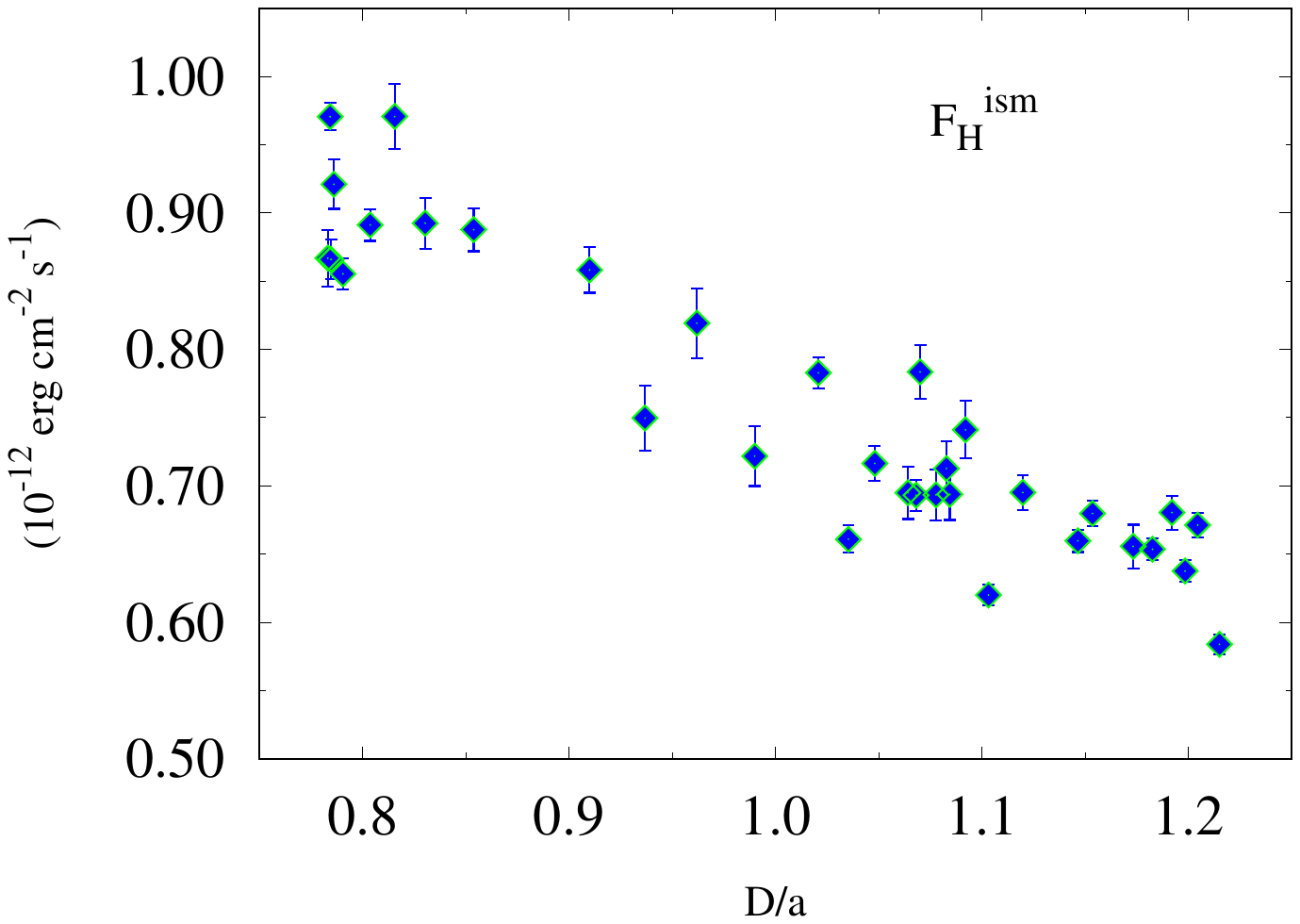}
\caption{Variation of the ism-corrected X-ray flux in 2.0-10.0 keV energy range ($F^{ism}_{H}$) from HD\,93250 obtained after X-ray spectral fitting as a function of the binary separation (D normalized to semi-major axis `a').
\label{fig:flux_bsep}}
\end{figure}

One of the modeled X-ray spectra of HD\,93250 is shown in Fig. \ref{fig:hd93250_spectra}. We show the phase-folded time series of best-fit parameters, $norm_1$, $norm_2$, $norm_3$,  and $N_H^{local}$, in Fig.,\ref{fig:hd93250_spec_par}. Even though we still notice some dispersion in the parameter values obtained at similar orbital phases, this constitutes our best set of results. The observed and ISM-corrected X-ray flux was also estimated in broad (F$_{B}^{obs,ism}$), soft (F$_{S}^{obs,ism}$), and hard (F$_{H}^{obs,ism}$) energy bands. The variation of X-ray flux as a function of the orbital phase is shown in Fig. \ref{fig:flux_phase} in different energy bands. The phase-locked X-ray flux variations are clearly displayed in this figure where it is maximum around the periastron and becomes minimum close to the apastron. Fig. \ref{fig:flux_bsep} displays the hard X-ray flux as a function of the varying binary separation (D/a; binary separation normalized with the semi-major axis `a' taken from \citealt{2017A&A...601A..34L}). 

\section{Discussion}\label{disc}
\subsection{X-ray luminosity}\label{Xlum}
The spectral classification of HD\,93250 is that of a pair of O4 stars, but the luminosity class is not fully established yet. Typical masses for O4 stars reported by \citet{2005A&A...436.1049M} are about 46, 49, and 58 M\,$\odot$, for dwarfs, giants, and supergiants, respectively. Given the total mass for the system proposed by \citet{2017A&A...601A..34L}, equal to 84 M\,$\odot$, it is less likely that the stars are supergiants. As a result, in compliance with the luminosity classes proposed for the stars in HD\,93250 (see Sect.\,\ref{intro}), let's assume that their luminosity class ranges between V and III. Adopted typical bolometric luminosities for O4 stars for both luminosity classes are quoted in Table\,\ref{adoptedpar}.

\begin{table}[h!]
\caption{Summary of the adopted parameters for the O4 components of HD\,93250 assuming V and III luminosity classes \citep{2005A&A...436.1049M,2012A&A...537A..37M}}.\label{adoptedpar}
\centering  
\begin{tabular}{l c c}
\hline
 & V & III \\
\hline
$L_{bol}$ (erg\,s$^{-1}$) & $1.84\,\times\,10^{39}$ & $2.54\,\times\,10^{39}$ \\
${\dot M}$ (M$_\odot$\,yr$^{-1}$) & $1.46\,\times\,10^{-6}$ & $2.88\,\times\,10^{-6}$ \\
$V_\infty$ (km\,s$^{-1}$) & 3600 & 2950 \\
$T_{eff}$ (K) & 43400 & 41500 \\
\hline
\end{tabular}
\end{table}

Assuming the distance to HD\,93250 is about 2370 pc \citep[Gaia DR3 catalog][]{2021AJ....161..147B}, fluxes corrected for interstellar absorption were converted into X-ray luminosities. For this discussion, let's consider the minimum and maximum flux values, obtained close to apastron ($f_{X,min} \sim 3.2\,\times\,10^{-12}$ erg\,cm$^{-2}$\,s$^{-1}$) and periastron ($f_{X,max} \sim 4.8\,\times\,10^{-12}$ erg\,cm$^{-2}$\,s$^{-1}$), respectively. This converts into $L_{X,min} \sim 2.15\,\times\,10^{33}$ and $L_{X,max} \sim 3.23\,\times\,10^{33}$ erg\,s$^{-1}$. We thus obtain ranges of $L_X/L_{bol}$ ratios (counting twice the individual $L_{bol}$) of (5.8 - 8.8\,)$\times\,10^{-7}$  and (4.2 - 6.4\,)$\times\,10^{-7}$, assuming dwarfs and giants, respectively. These values are significantly above the usual $L_X/L_{bol}$ ratio for single O stars of about 10$^{-7}$, valid for stars not too close to the transition to the WN type \citep{2013MNRAS.429.3379O,2013NewA...25....7D}. This X-ray luminosity excess is attributable to the thermal X-ray emission arising from the colliding-wind region (CWR), on top of the emission from individual stellar winds.

Energy budget considerations about the wind kinetic power,
\begin{equation}\label{pkin}
P_{kin} = \frac{1}{2}\,{\dot M}\,v_\infty^2 = 3.155\,\times\,10^{35}\,{\dot M}_{-6}\,v_{\infty,8}^2\,\,\,\,\mathrm{(erg\,s^{-1})}
\end{equation}
\noindent where ${\dot M}_{-6}$ is the mass loss rate in units of 10$^{-6}$\,M$_\odot$\,yr$^{-1}$, and $v_{\infty,8}$ is the terminal velocity in units of 10$^8$\,cm\,s$^{-1}$, also deserve to be addressed. Based on values quoted in Table\,\ref{adoptedpar} for wind parameters, we derived a total $P_{kin}$ values of $1.2\,\times\,10^{37}$ and $1.58\,\times\,10^{37}$ erg\,s$^{-1}$ for the sum of both winds, assuming classes V and III, respectively. Quantifying the fraction of mechanical energy effectively converted into X-ray emission in the CWR requires estimating $L_{X,CWR}$, which is the contribution of the X-ray luminosity emerging from the system that is coming from the CWR. The measured $L_{X}$ has to be corrected for the contributions from the two stellar winds, that we assume to be $10^{-7}\,L_{bol}$, 
\begin{equation}\label{LXcwr}
L_{X,CWR} = L_{X} - 2\,\times\,10^{-7}\,L_{bol}\,\,\,\,\mathrm{(erg\,s^{-1}).}
\end{equation}
Using Eq.\,\ref{LXcwr}, and considering the minimum and maximum values of $L_{X}$, we obtain ranges of $L_{X,CWR}$ of (1.78 - 2.86)\,$\times\,10^{33}$ and (1.64 - 2.72)\,$\times\,10^{33}$ erg\,s$^{-1}$, respectively for both considered luminosity classes. These numbers convert to $L_{X,CWR}/P_{kin}$ ratios of (1.5 - 2.4)\,$\times\,10^{-4}$ and (1.0 - 1.7)\,$\times\,10^{-4}$. These ratios are in agreement with other O-type systems with periods of several months or years \citep{2015MNRAS.451.1070D}.

\subsection{X-ray variability}\label{Xvar}

The normalization parameter (Fig.\,\ref{fig:hd93250_spec_par}) of the softest emission components doesn't display any specific trend. The dispersion is quite significant, with no noticeable increase close to periastron. We have to notice that the soft component is undoubtedly the one that is reproducing most of the soft X-ray emission from individual stellar winds (as a result of lower pre-shock velocity). The relative contribution of the CWR emission is the weakest in that component, as compared to the other two emission components. On the opposite, the hardest component is the one that displays the clearest trend, with a significant and gradual increase of $norm_3$ between apastron and periastron, reaching a peak at the latter orbital phase. The reverse trend is suggested when moving from periastron to apastron, but the time sampling of that half of the orbit is sparser. Given that the harder part of the spectrum is very likely dominated by the CWR emission (as a result of the greater pre-shock velocity as compared to intrinsic wind shocks), the discussion of the binary modulation should focus on that part. Regarding the evolution of the local hydrogen column, Fig.\,\ref{fig:hd93250_spec_par} suggests a peak between phases 0.5 and 1.0, thus not coincident with periastron. This may be explained by the orientation of the system that favors some enhanced local absorption when one of the stars is in front of a significant part of the CWR. However, one has to be cautious as the local absorption is sensitive to the softer part of the spectrum that seems to be rather poorly constrained given the lack of a clear trend displayed by the normalization of the soft component. Apart from a very few deviant points, the trend displayed by the $norm_2$ parameter is similar to that of $norm_3$, but with a somewhat lower peak-to-bottom amplitude. That component, with a plasma temperature of about 1.0 keV, is significantly due to the CWR but still with some likely contribution from the individual winds.

As illustrated by both count rates and fluxes, the X-ray emission is significantly varying as a function of the orbital phase, in full agreement with the ephemeris published by \citet{2017A&A...601A..34L}. Given the wind parameters of the components of the system, one can estimate the cooling parameter ($\chi$) as defined by \citep{1992ApJ...386..265S} to establish the nature of the shock (adiabatic or radiative) in the CWR as 

\begin{equation}
 \chi = \frac{v^{4} D}{\dot{M}}
\end{equation}

\noindent
 where $v$ is the pre-shock wind velocity in 1000 km\,s$^{-1}$ units, $D$ is the distance from the star to the shock in $10^{7}$ km and $\dot{M}$ is the mass loss rate in 10$^{-7}$ M$_{\odot}$\,yr$^{-1}$. A $\chi$ value well above 1 indicates a full adiabatic regime, implying that the X-ray emission from the CWR should follow a $1/D$ trend, where $D$ is the stellar separation in the system. According to the parameters quoted in Table\,\ref{adoptedpar}, for a stagnation point located midway between the two stars and a separation based on the astrometric orbit proposed by \citet{2017A&A...601A..34L}, we determine that $\chi$ varies in the ranges 520-813 and 119-185 assuming O4\,V and O4\,III components, respectively. In these ranges, the lower (upper) value corresponds to periastron (apastron). The estimated value is much greater than 1, warranting a full adiabatic regime. The $1/D$ variation of F$_{H}^{ism}$ is clearly manifested in Fig.\,\ref{fig:flux_bsep}. 

When comparing apastron to periastron, the stellar separation changes by a factor $(1 + e)/(1 - e)$, where $e$ is the eccentricity. With an $e = 0.22$, the separation ratio is of the order 1.6. Besides, the ratio of maximum and minimum values of $L_{X,CWR}$ given in Sect.\,\ref{Xlum} are 1.61 and 1.66 for V and III luminosity classes, respectively. These values are similar to the maximum-to-minimum count rate ratio obtained in the hard band (see Sect.\,\ref{lc}) and flux ratio from spectral analysis in hard band (=1.67). One can also characterize the amplitude of the variation of the emission from the colliding winds through the periastron-to-apastron $norm_3$ ratio as this parameter scales with the emission measure of the hottest component, and we obtain about 1.5. These values are in quite fair agreement with the expected separation ratio. It is interesting to note that based on our measured ratios ($r$), we would estimate the eccentricity $e = (r - 1)/(r + 1)$ of the system to be in the 0.20 - 0.25 range, demonstrating the relevance of this approach to estimate some basic orbital parameters even in the absence of fully independent orbital solution.

Regarding the local absorption by the wind material, we caution that the use of a slab-geometry model such as the one used in this study leads to limitations in the interpretation. Ideally, a full hydro-radiative simulation of the binary system should more consistently account for this effect, but this is fully out of the scope of this study. If the target was a single star wind, the ideal approach to account for local wind absorption would the one developed by \citet{Leutenegger}. However, given the binary nature of the system (including stellar winds and their collision), no appropriate tool is at our disposal. The simple modeling by a slab-geometry model is however enough to achieve a qualitative description of the overall behavior of the system, especially in terms of the orbital modulation of the thermal X-ray flux.

\subsection{HD\,93250 in hard X-rays}
{\it NuSTAR} data provides an opportunity to investigate hard X-rays from HD\,93250. In the context of colliding-wind binaries, hard X-ray emission could arise from inverse Compton (IC) scattering of photospheric photons by relativistic electrons accelerated by shocks in the colliding-wind region. This has been confirmed in the case of two massive binaries: $\eta$\,Car \citep{2018NatAs...2..731H} and Apep \citep{2023A&A...672A.109D}. Given that HD\,93250 is known to be a particle accelerator thanks to its non-thermal radio emission (see Sect.\,\ref{radio}), it is worth considering its potential non-thermal emission in X-rays that requires investigating photon energies above 10 keV to avoid severe contamination by thermal X-rays \citep{2007A&ARv..14..171D}. 

\begin{figure}[h!]
\centering
  \includegraphics[width=1.0\columnwidth,trim={0.0cm 0.0cm 4.0cm 1.0cm}]{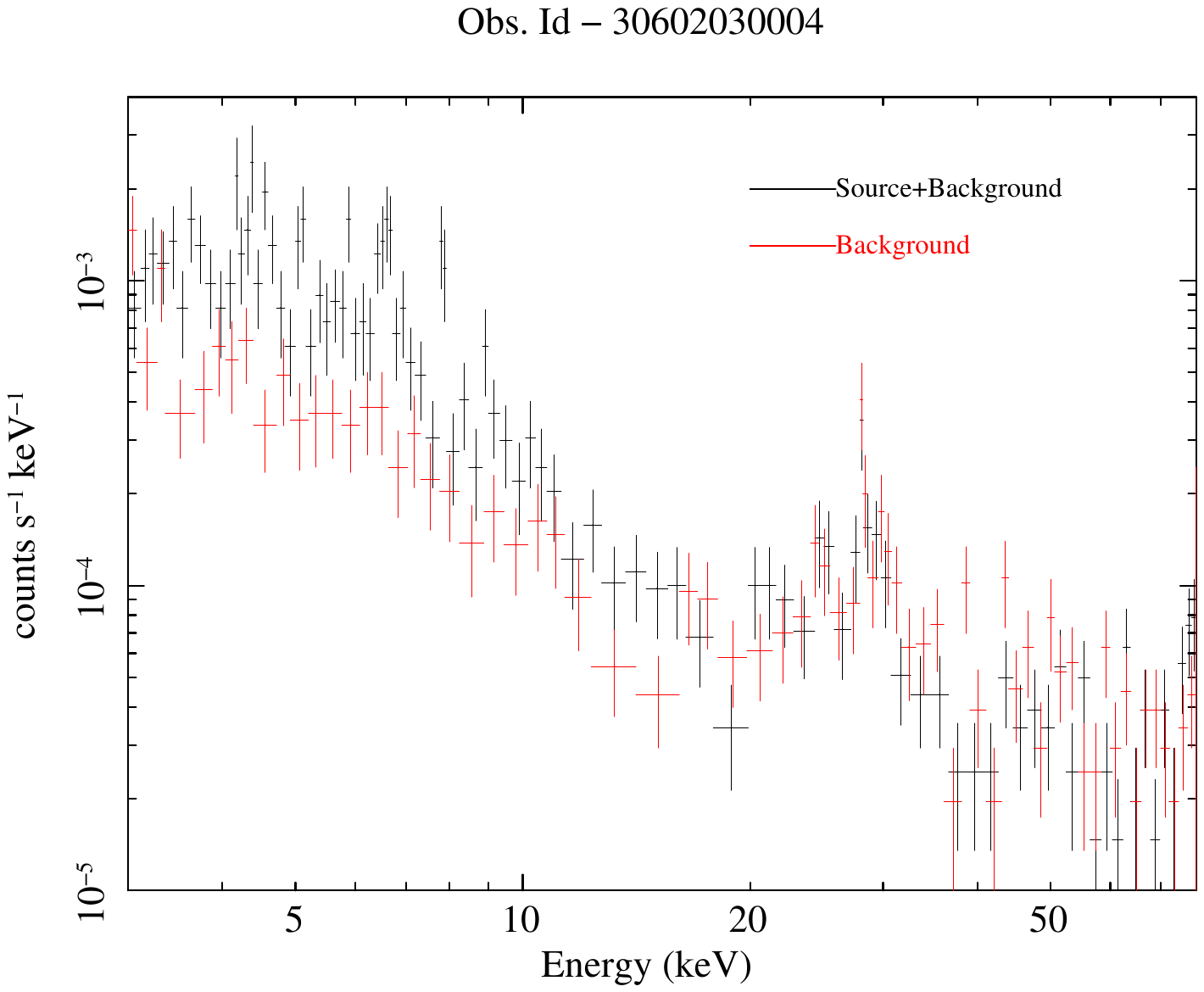}
\caption{\textit{NuSTAR}-FPMA spectra extracted at the position of HD\,93250 and from the background region as indicated in Fig. \ref{Nustar_image}.
\label{nustar_spectra}}
\end{figure}

HD\,93250 appears as a weak point source on the {\it NuSTAR} image (Fig.\,\ref{Nustar_image}). Figure\,\ref{nustar_spectra} shows spectra extracted in the source and background regions. While some excess emission attributable to thermal emission from the source is measured in the soft part of the spectrum, no significant excess emission emission is detected above 10 keV. The source counts in 10.0--78.0 keV for all of the epochs of {\it NuSTAR} observation given in Table \ref{nustar_table} were of the order of the background counts. We can thus claim that no non-thermal emission is detected. 

We determined upper limits on the hard-X-ray emission using the same approach as \citet{2019MNRAS.487.2624A}. To do so, we selected observation ID 30402001004, as it is the closest to periastron passage expected to coincide with a maximum in IC emission (proportional to the local radiative energy density). First, we measured the number of counts in the source region above 10 keV. According to \citet{2013ApJ...770..103H} a 40 arcseconds extraction radius corresponds to an encircled energy fraction of 60$\%$, and at an offset of 8 arcminutes, the effective area is about 25$\%$ of the maximum. Applying these two corrections yields a corrected count number, $C_{cor}$. We determined a count threshold ($C_{max}$) corresponding to a logarithmic likelihood ($L$) of 12, translating into a probability ($P$) to find a count number in excess of $C_{max}$ of about $6\,\times\,10^{-6}$ ($L = -ln P$), under the null hypothesis of pure background Poisson fluctuations. We then divided the count excess ($C_{max} - C_{cor}$) by the effective exposure time to determine an upper limit on the count rate. Our results are summarized in Table\,\,\ref{uplim}. 

\begin{table}
	\caption{Upper limits on the count rate from \textit{NuSTAR} data.\label{uplim}}
	\begin{center}
		\begin{tabular}{l c c}
			\hline
			& FPMA & FPMB  \\
			\hline
			\vspace*{-0.2cm}\\
			$C$ (cnt)& 143 & 122 \\
			$C_{cor}$ (cnt) & 954 & 814 \\
			$C_{max}$ (cnt) & 1092 & 942 \\
			$C_{max} - C_{cor}$ (cnt) & 138 & 128 \\
			$CR$ (cnt\,s$^{-1}$) & 4.2\,$\times$\,10$^{-3}$ & 3.3\,$\times$\,10$^{-3}$ \\
			\vspace*{-0.2cm}\\
			\hline
		\end{tabular}
	\end{center}
\end{table}

Based on these count rate upper limits, we used the WebPIMMS on-line tool\footnote{\url https://heasarc.gsfc.nasa.gov/cgi-bin/Tools/w3pimms/w3pimms.pl} to estimate the corresponding flux, assuming a power law emission with a photon index equal to 1.5. The latter value is expected for relativistic electrons accelerated through Diffusive Shock Acceleration in the test-particle regime by adiabatic high Mach number shocks. We obtain an upper limit on the flux between 0.1 and 100 keV of the order of $1.5\,\times\,10^{-12}$\,erg\,cm$^{-2}$\,s$^{-1}$, which converts into a luminosity of the order of 10$^{33}$\,erg\,s$^{-1}$. This corresponds to a fraction of the order of 10$^{-4}$ of the total wind kinetic power of the system (see Sect.\,\ref{Xlum}). As a comparison, \citet{2023A&A...672A.109D} report on a conversion efficiency of kinetic power to IC scattering of about $1.5\,\times\,10^{-4}$ in the case of Apep, that is a bit above our limit for HD\,93250. Given the sensitivity of {\it NuSTAR}, provided our target is indeed producing non-thermal X-rays, it is injecting too little energy in non-thermal processes to warrant a detection. This is fully in agreement with the idea that non-thermal X-ray emitters have a chance to be revealed provided their input energy reservoir, which is their wind kinetic power, is abundant enough. This is the main reason behind the very few detections reported so far.

\subsection{Considerations about the radio emission from HD\,93250}\label{radio}

Besides the thermal X-ray emission from the colliding winds and considerations about non-thermal X-rays, it is relevant to consider the question of the synchrotron radio emission associated with shocks in the CWR of tens of massive binaries \citep{2013A&A...558A..28D,2017A&A...600A..47D}. In this context, the synchrotron radio emission is also seen as a signature of colliding winds.

In order to investigate the nature of the radio emission from HD\,93250, \citet{1995ApJ...450..289L} observed it at 8.64 and 4.80 GHz frequencies. The source was detected only at 8.64 GHz with a flux density of 1.36 $\pm$ 0.17 mJy. Assuming this radio emission was only due to a single star wind, the subsequent derivation of the associated mass-loss rate pointed toward an excessive value, suggesting the measured radio emission was not only due to thermal free-free emission from a single star wind. As this was likely attributable to a non-thermal contribution in the spectrum, this led HD\,93250 to be included in the catalog of particle accelerating colliding wind binaries \citep[PACWBs,][]{2013A&A...558A..28D}, although it was tagged with a low confirmation level flag. Besides the X-ray emission from the CWR investigated in this paper, this potential synchrotron radio emission constitutes a second aspect of the shock physics associated with its binarity. 

In light of the revision of the properties of HD\,93250 since the publication by \citet{1995ApJ...450..289L}, it is certainly relevant to reconsider the question of its radio emission. We estimated the free-free thermal radio emission from the winds of the components of HD\,93250 following the approach described by \citet{2018A&A...620A.144D} (see their Appendix B), based on the theoretical development established by \citet{1975MNRAS.170...41W}. In agreement with the prescription we applied in Sect.\,\ref{Xlum}, let's consider both dwarf and giant luminosity classes. We used the parameters quoted in Table\,\ref{adoptedpar} along with a clumping factor of 4 whose main effect is to enhance the thermal free-free emission for a given value of the mass loss rate (not considered at the time of the radio measurement). As for the X-ray emission, the putative synchrotron radio emission should emerge from the CWR, on top of the thermal free-free emission from the two winds. Adding together the predicted flux density from the two stellar winds in HD\,93250, assuming main-sequence stars we obtain about 0.02 mJy at 4.8 GHz and 0.03 mJy at 8.6 GHz. Assuming alternatively giant stars, we obtain about 0.08 mJy at 4.8 GHz and 0.11 mJy at 8.6 GHz. We can thus conclude that the measurement at 8.6 GHz made by \citet{1995ApJ...450..289L} is at least one order of magnitude brighter than our prediction for pure thermal emission. As uncertainties on the selected parameters are unlikely to lead to such a large discrepancy in our prediction, we confirm that the unique radio measurement of the system is pointing to a likely PACWB status.

At the time of this unique radio measurement (7 - 9 September 1994), according to the ephemeris of \citet{2017A&A...601A..34L} to orbital phase was almost coincident with periastron. This may explain a significant attenuation of the synchrotron emission due to free-free absorption by the stellar wind material (more pronounced at 4.80 GHz than at 8.64 GHz). The case of HD\,93250 certainly deserves additional radio measurements to clarify its behavior.

\section{Summary and conclusions}\label{conc}

We report on a detailed analysis of the O4 + O4 binary system HD\,93250 in soft and hard X-rays. In light of the characterization of the astrometric orbit of the system, all quality archived data sets obtained with {\it XMM-Newton} were used, summing up 33 epochs spread over about 19 years. Time series analysis based on count rates reveals a period of 193.8 $\pm$ 1.3 d, in full agreement with the orbital period determined by \citet{2017A&A...601A..34L}. This suggests that a significant fraction of X-rays measured in the EPIC bandpass arise from the colliding-wind region. This statement is notably reinforced by the morphology of the spectrum that is moderately hard, as expected from a plasma heated by colliding wind at speeds of a few thousand of km\,s$^{-1}$.

The spectral analysis reveals an X-ray spectrum compatible with optically thin thermal emission by a plasma with a temperature distribution ranging from a few 10$^6$\,K to several 10$^7$\,K. Our best results were obtained using an absorbed (interstellar and local) three-temperature model. Considering the need to perform physically consistent modeling, we explored the parameter space to converge a set of valid parameters with the strong requirement to display a smooth evolution as a function of the orbital phase. The best physically valid set of solutions was obtained by freezing the plasma temperatures, allowing the normalization parameters and $N_H^{local}$ to vary. 

Depending on the assumption of the luminosity class of the stars (V or III), the over-luminosity factor (as compared to the expected emission from non-interacting winds) ranges between 3 and 7, in agreement once again with a significant contribution of the X-ray emission from the CWR. We also estimate that the fraction of the wind kinetic power converted into thermal X-rays in the CWR is of the order of a few times 10$^{-4}$, in agreement with other O-type systems with periods of at least several months. A lack of detection of HD\,93250 above 10 keV based on {\it NuSTAR} data has been reported, and we determined an upper limit on the putative inverse Compton scattering emission that is about $10^{-4}$ times the wind kinetic power in the system.

We re-evaluated the question of the Particle-Accelerating Colliding-Wind Binary status of HD\,93250. Based on the expected wind parameters, the calculated expected thermal free-free emission from individual stellar winds is about one order of magnitude too low to explain the unique existing radio measurement, lending support to the idea it is indeed a synchrotron radio emitter. However, a unique measurement is not enough to characterize its behavior. Besides the thermal X-ray emission investigated in detail in this paper, the non-thermal emission in the radio domain constitutes a second aspect of shock physics in colliding-wind binaries that still needs to be tackled for this system.

Finally, we stress that X-ray times series constitute relevant independent tools to identify massive binaries, determine their orbital period and estimate their eccentricity, provided the sampling of the orbit is good enough. This is relevant in terms of the efficiency of indirect techniques, such as X-ray time analysis, to investigate the multiplicity of massive stars and extract relevant information on their basic orbital parameters independently of more classical methods based on radial velocity curves or astrometric monitoring. 

\begin{acknowledgements}
  We thank the referee for the careful reading of the manuscript and giving us constructive comments and suggestions. BA acknowledges the grant provided by Wallonia Brussels International to carry out this work. This research has made use of observations obtained with \textit{XMM$-$Newton}, an ESA science mission with instruments and contributions directly funded by ESA Member States and NASA. Additionally, \textit{NuSTAR} data has been utilized which is a project led by the California Institute of Technology, managed by the Jet Propulsion Laboratory, and funded by the National Aeronautics and Space Administration.
\end{acknowledgements}

%
%

\bibliographystyle{aa}
\bibliography{ref2}{}

\end{document}